\documentclass[12pt,preprint]{aastex}

\def\arcs{\char'175\ }
\def\arcsc{\char'175 }
\def\arcm{$^\prime$}
\def\arcmn{$^\prime$}

\def\etal{et~al.\ }

\def\hub{\ifmmode H_\circ\else H$_\circ$\fi}

\newbox\grsign \setbox\grsign=\hbox{$>$} \newdimen\grdimen \grdimen=\ht\grsign
\newbox\simlessbox \newbox\simgreatbox
\setbox\simgreatbox=\hbox{\raise.5ex\hbox{$>$}\llap
     {\lower.5ex\hbox{$\sim$}}}\ht1=\grdimen\dp1=0pt
\setbox\simlessbox=\hbox{\raise.5ex\hbox{$<$}\llap
     {\lower.5ex\hbox{$\sim$}}}\ht2=\grdimen\dp2=0pt

\newbox\simppropto
\setbox\simppropto=\hbox{\raise.5ex\hbox{$\sim$}\llap
     {\lower.5ex\hbox{$\propto$}}}\ht2=\grdimen\dp2=0pt

\begin{document}

\title{A Library of Integrated Spectra of Galactic Globular Clusters}

\author{Ricardo P. Schiavon}
\affil{Astronomy Department, University of Virginia,  P O Box 3818,
    Charlottesville, VA 22903, USA}
\email{ripisc@virginia.edu}

\author{James A. Rose}
\affil{Department of Physics \& Astronomy, University of North Carolina,
    Chapel Hill, NC 27599, USA}
\email{jim@physics.unc.edu}

\author{St\'ephane Courteau}
\affil{Department of Physics, Queen's University, Kingston, ON K7L 3N6,
Canada}
\email{courteau@astro.queensu.ca}

\and

\author{Lauren A. MacArthur}
\affil{Department of Physics \& Astronomy, University of British Columbia,
6224 Agricultural Road, Vancouver, BC Canada, V6T 1Z1}
\email{lauren@physics.ubc.ca}

\begin{abstract}
We present a new library of integrated spectra of 40 Galactic globular
clusters, obtained with the Blanco 4-m telescope and the R-C spectrograph
at the {\it Cerro Tololo Interamerican Observatory}.  The spectra cover
the range $\sim$ 3350 -- 6430 ${\rm\AA}$ with $\sim$ 3.1 ${\rm\AA}$
(FWHM) resolution. The spectroscopic observations and data reduction
were designed to integrate the full projected area within the cluster
core radii in order to properly sample the light from stars in all
relevant evolutionary stages. The S/N values of the flux-calibrated
spectra range from 50 to 240/${\rm\AA}$ at 4000 ${\rm\AA}$ and from 125
to 500/${\rm\AA}$ at 5000 ${\rm\AA}$. The selected targets span a wide
range of cluster parameters, including metallicity, horizontal-branch
morphology, Galactic coordinates, Galactocentric distance, and
concentration. The total sample is thus fairly representative of the
entire Galactic globular cluster population and should be valuable
for comparison with similar integrated spectra of unresolved stellar
populations in remote systems. For most of the library clusters, our
spectra can be coupled with deep color-magnitude diagrams and reliable
metal abundances from the literature to enable the calibration of stellar
population synthesis models.  In this paper we present a detailed account
of the observations and data reduction.  The spectral library is publicly
available in electronic format from the National Optical Astronomical
Observatory website.

\end{abstract}


\section{Introduction}

The determination of the stellar content of unresolved systems from
observations of their integrated light is a fundamental challenge to
extragalactic astronomers. The problem has been traditionally approached
by comparing the integrated properties (spectral energy distributions,
line strengths, colors) of remote systems with those of better-understood
local templates (e.g. Faber 1973, Burstein \etal 1984, Rose 1985). More
recently, stellar population synthesis (SPS) models combined with our
knowledge of the stellar evolution of stars and of the behaviour of their
emitted light as a function of fundamental stellar parameters have been
constructed to predict integrated properties of single and composite
stellar populations in integrated light (e.g. Le Borgne \etal 2004,
Bruzual \& Charlot 2003, Thomas, Maraston \& Bender 2003, Vazdekis 1999,
Worthey 1994).

In this context, the Galactic globular cluster (GC) system plays a
crucial role. It consists of a set of increasingly well studied stellar
populations which are close enough that their constituent stars can
be resolved and studied individually and meaningful color-magnitude
diagrams can be constructed, and yet their compact structures enable
observations of their integrated light relatively easily. Integrated
observations of GCs can be used to construct calibrations of integrated
observables as a function of the parameters that characterize stellar
populations (mainly age, metallicity, and abundance pattern). Such
calibrations provide useful, fairly model-independent tools for the
conversion of integrated observables of distant systems into fundamental
stellar population parameters. Moreover, comparisons of SPS models
with integrated observations of Galactic GCs has become commonplace
(e.g. Proctor, Forbes \& Beasley 2004, Schiavon \etal 2004a,b, Maraston
\etal 2003, Puzia \etal 2002, Schiavon \etal 2002a,b, Beasley, Hoyle
\& Sharples 2002, Vazdekis \etal 2001, Gibson \etal 1999, Schiavon \&
Barbuy 1999, Bruzual \etal 1997, Rose 1994, and references therein). Such
fundamental comparisons aim primarily at verifying the validity of the
models, whose predictions must match the integrated properties of GCs
for the known set of input stellar population parameters.

Reliable databases of integrated properties of Galactic GCs are therefore
precious resources and numerous efforts have been invested towards their
collection in the past. Notable among these are the works of Puzia
\etal (2002), Cohen, Blakeslee \& Ryzhov (1998), Covino, Galletti \&
Pasinetti (1995), Bica \& Alloin (1986), Burstein \etal (1984), and Zinn
\& West (1984), where integrated spectra of moderate to large-sized
samples of Galactic GCs were collected. In spite of these and other
previous works, there remains considerable room for improvement to the
available spectroscopic databases which are deficient in one or more of
such crucial features as size, a representative range of metallicities
and horizontal branch (HB) morphologies, signal-to-noise ratio (S/N),
resolution, and homogeneity. These features are all essential both for
a robust calibration of stellar population synthesis models and for the
direct comparison with observations of distant systems. This work aims
at providing this much needed improvement.

We present new, high-S/N, moderately high-resolution, integrated spectra
of 40 Galactic globular clusters obtained with the same instrumental
setup and over a single observing run. This is the largest collection of
optical integrated spectra of Galactic GCs to date with such high S/N
and resolution, and should serve as a prime database for comparison
with observations of remote stellar systems and the calibration of
SPS models. This paper focuses on the detailed description of our
observations and data reduction procedures, and the presentation of
the final data products. A first exploration of this spectral library,
focussed on the effects of HB morphology on different Balmer lines,
was presented in Schiavon \etal (2004b).

The paper is organized as follows: we describe the sample selection
and observations, in Sections \ref{sec:sselection} and \ref{sec:obsdata},
respectively, and the reductions are discussed in Section
\ref{sec:reductions}. The final spectra are presented in Section 
\ref{sec:spectra} and Section \ref{sec:notes} contains notes on 
individual clusters. Our results are summarized in Section 
\ref{sec:summary}.

\section{Sample Selection} \label{sec:sselection}

\begin{figure}
\plotone{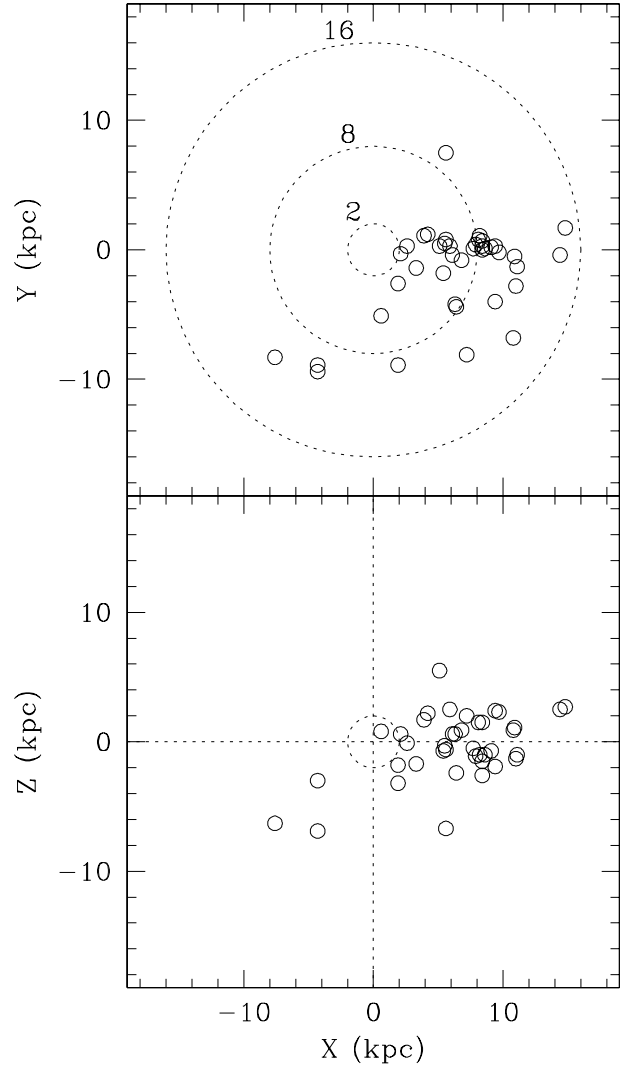}
\caption{Spatial distribution of the sample clusters.}
\label{fig:spdist}
\end{figure}

Our choice of GC targets was chiefly aimed at producing a spectral
library that is representative of the Galactic GC system. Therefore,
GCs spanning a wide range of metallicities, HB morphologies, and
Galactocentric distances were chosen. Another important consideration is
that the spectral library will become more useful if GC characteristic
parameters such as age, metallicity, and HB morphology are known
accurately. Therefore, we gave priority to GCs for which [Fe/H]
values, based on abundance analyses of high resolution spectra of
member stars (e.g. Carretta \& Gratton 1997, Kraft \& Ivans 2003),
are available. For the same reason, we also preferred GCs for which
high-quality color-magnitude diagrams, preferrably reaching below the
turn-off, are available from the literature (e.g. Piotto \etal 2002).
The combination of such data with our spectral library optimizes the
construction of reliable calibrations of integrated GC properties with
fundamental determinations of basic parameters like age, metallicity and
HB morphology. The distribution of our GC sample in Galactic coordinates
is shown in Figure~\ref{fig:spdist}.


The main characteristics of our globular cluster sample are summarized
in Table~1.  We list in columns (1) and (2) the cluster IDs, in columns
(3) and (4) the positions in galactic coordinates, and in column (5)
Galactocentric distances in kpc. In column (6) we list core radii, $r_c$,
in arcminutes, in column (7) the central concentration parameter $c =
\log{(r_t/r_c)}$, where $r_t$ is the tidal radius, and in column (8)
reddening values, $E(B-V)$, are listed. We list in column (9) [Fe/H]
values from high-resolution abundance analyses, when available, and
those determined by Schiavon \etal (2005, in preparation) otherwise. In
column (10) the Lee, Demarque \& Zinn (1994) parameter describing the
morphology of the HB is listed. Finally, in column (11) we provide a
non-exhaustive list of recent references where deep color-magnitude
diagrams for the sample clusters can be found. The data in columns
(3), (4), (5), and (8) were taken primarily from Piotto \etal (2002),
or otherwise from the compilations by Djorgovski \& Meylan (1993),
Djorgovski (1993), and Peterson (1993). The [Fe/H] values from Schiavon
\etal (2005, in preparation) are derived from a new metallicity scale
for Galactic GCs, based on measurements of iron absorption lines
and state-of-the-art high-resolution abundance analyses of cluster
members.  The HB morphologies listed in column (10) were taken from
Borkova \& Marsakov (2000). For GCs whose HB morphologies are not
given by Borkova \& Marsakov, we provide qualitative estimates from
inspection of the color-magnitude diagrams of Piotto \etal (2002). Core
radii and concentration parameters were taken from Harris (1996)
({\tt http://physwww.mcmaster.ca/$^\sim$harris/mwgc.dat}).

\section{Observational Data} \label{sec:obsdata}

\subsection{Instrumental Setup} \label{sec:setup}

All observations were obtained with the Ritchey-Chretien spectrograph
mounted on the 4-m Blanco telescope at the {\it Cerro Tololo
Inter-American Observatory} (CTIO) on the nights of 24-27 April 2003.
A Loral 3K x 1K CCD detector was used with an Arcon controller.
The grating KPGL\#1, with 632 grooves/mm, was used in first order with
the blue collimator, yielding a dispersion of 1.0 \AA/pix over the
wavelength range 3360--6430 \AA.  A spectral resolution of 3.1 \AA \
FWHM was achieved in the central $\sim$1000 \AA \ of the dispersion
curve, deteriorating to $\sim$3.6 \AA \ FWHM at the blue and red ends
of the spectrum. A cubic polynomial fit to the FWHM of arc lines as a
function of wavelength produced the following curve:

\begin{equation}
FWHM = 15.290 - 6.079 \times 10^{-3} \lambda + 9.472 \times 10^{-7} \lambda^2
- 4.395 \times 10^{-11} \lambda^3,
\end{equation}

\noindent where $\lambda$ is the wavelength. Both FWHM and $\lambda$
are given in \AA.

The spectra were binned by a factor of two along the slit, yielding
a spatial scale of 1\arcs/pixel. The gain of the CCD was set to 1.0
e$^{-}$/ADU, and the read noise was determined to be 7.3 e${^-}$.
For all observations, except those of flux standards, the slit width
was set at 225 $\micron$ or 1\farcs5.  However, as we discuss further in
\S\ref{sec:anomalies}, there were problems with the slit width in the
second half of the last night of observations, due to a failure in the
slit motor and its subsequent reinitialization.  The change in slit width
also produced a change in spectral resolution for those observations.
For the first two nights, the $\sim$ 5\farcm5-long slit was oriented
in the North-South direction, and was along the East-West direction for
the final two nights.

\subsection{Calibration Exposures}

At the beginning of each night a sequence of 25 bias frames and
25 dome flat field frames were acquired.  The bias level fluctuated
throughout the observing run, but could always be subtracted adequately
(see \S\ref{sec:anomalies}). Twilight exposures were recorded on the
first two nights to provide illumination correction along the slit. The
combination of a set of 5 arc lamp exposures taken at the beginning of
each night provided a high-S/N master arc. Finally, on the last night,
we observed the G star HD76151 at several positions along the slit to
aid in mapping optical distortions in the spectrograph.  Specifically, we
observed HD76151 at the center of the slit, and repeated the observation
with the slit offset by 1\arcmn, 2\arcmn, and 2\farcm5 East and West of
the central position.

\subsection{Globular Cluster Observations}

Due to the spatially extended nature of globular clusters, and our desire
to obtain their representative integrated spectra, each globular cluster
observation was obtained by drifting the spectrograph slit across the
core diameter of the cluster. The cluster core radii are listed in 
Table 1. 
The telescope was positioned so as to offset the slit
from the cluster center by one core radius. A suitable trail rate would
then allow the slit to drift across the cluster core diameter during
the typically 15 minute-long exposure.

In addition to the trail across the cluster core, a suitable background
determination is required. We considered the extraction of a background
spectrum from the ends of the $\sim$5\farcm5 slit.  However there are
several disadvantages to this procedure. For many of the clusters,
there would still be significant contamination from the cluster itself
even at a radial distance of $\sim$2--3\arcmn.  Use of the extreme
ends of the slit for sky subtraction also requires a very secure
mapping of the slit illumination and of the geometric distortions
in the spectrograph.  As well, there is the danger of deteriorating
focus at large distances from the slit center. All of these problems
are circumvented by taking a separate sky exposure from the cluster
exposure, and then directly subtracting the sky frame from the cluster
frame\footnote{This procedure is of course only recommended when the
sky background does not fluctuate significantly in timescales comparable
to the exposure times, a condition that is met in the optical, but not,
for instance, in the near-infrared.}. This approach also allows one to
carefully select patches of background that appear to be well matched
to that of the cluster, which is particularly important in regions
towards the Galactic bulge, where the background variations are large
due to non-uniform extinction and/or diffuse HII emission. We selected
background areas in the vicinity of each cluster from film copies of the
{\it European Southern Observatory/SRC} plates and/or print copies of
the {\it Palomar Observatory Sky Survey}. In crowded regions towards the
Galactic bulge, areas of typically 5\arcm\ $\times$ 5\arcmn\ or 10\arcm\
$\times$ 10\arcmn\ were selected for background scans.  The criteria
used in defining the scan region included visually similar level of
faint background stars as for the cluster, avoidance of discrete bright
stars, and, if applicable, similar level of HII emission as the cluster
(as best seen on the red sky survey print/film).

The typical observation sequence included an arc spectrum, a 15
minute trailed cluster observation, and a 15 minute trailed background
observation. For clusters with a low star background, the slit was fixed
at a particular sky position, i.e., no trail was made for the background
spectrum.  In the case of one cluster, NGC 6352, for which the background
spectrum was ``contaminated'' by the presence of several very bright B and
A stars that were covered in the trailed background spectrum, we found it
safer to use the ends of the slit from the cluster observation to extract
the background spectrum.  A journal of observations, providing complete
information regarding the globular cluster and background spectra,
is given in Table~2. 
We give in columns (1), (2), and (3)
the cluster IDs, the UT date of observation, and the exposure time (in
minutes), respectively.  The orientation of the spectrograph slit is
listed in column (4).  In column (5) we specify the angular distance,
in arcseconds, over which the spectrograph slit was trailed. A trail
of 50\arcs indicates that we trailed by $\pm$25\arcs on either side of
the cluster center.  We list in columns (6) and (7) the exposure time
and angular offset (from the cluster center), in arcminutes, for the
sky background exposure.  In column (8) we give the angular distance,
in arcminutes, over which we trailed the spectrograph slit for the sky
background exposure. A trail of 10\arcm\ indicates that we moved the
telescope by $\pm$5\arcm on either side of the sky offset position
given in column (7).  Finally, in column (9) we give the region extracted
along the slit for the given cluster.  Normally this was a symmetrical
extraction of $\pm$$r_c$ on either side of the cluster center, but in
some instances it was necessary to consider an asymmetrical extraction.

\subsection{Standard Star Observations}

In addition to the globular cluster observations, typically 2
or 3 spectrophotometric flux standards were observed per night.
For these observations the slit was widened to 10\arcsc to avoid
wavelength-dependent light losses through differential atmospheric
refraction. For the cluster observations the slit was trailed across
the cluster so that atmospheric dispersion is not an issue.

On each night several Lick/IDS standard stars were also observed to
transform our observations into the Lick spectral index system (Worthey
\etal 1994). To further tie in our observations with population synthesis
models, we also observed several stars from two more recent stellar
spectral libraries based on the coud\'e feed spectrograph at the Kitt
Peak National Observatory.  These libraries are both available at
the NOAO website.  We note here the more extensive current library,
called the Indo-US Library of Coud\'e Feed Stellar Spectra, which
is summarized in Valdes \etal (2004), and available at the {\tt URL:
http://www.noao.edu/cflib/}.

\subsection{Weather and Instrumental Anomalies} \label{sec:anomalies}

Since the purpose of this database is to provide a homogeneous set of
integrated spectra of Galactic globular clusters, we summarize here
all aspects of the observations that could generate a departure from a
strictly homogeneous dataset. We stress that in all cases these departures
are minor and can be appropriately corrected or accounted for.

\subsubsection{Bias Level Variations} \label{sec:bias}

From frequent monitoring of bias frames, we determined that the bias
was subject to large variations during the night.  The mean overscan
level fluctuated by up to 80 ADU (hence 80~e$^-$) during the night
and could jump by as much as 10 ADU between consecutive bias frames.
In addition, the shape of the overscan varied from one bias frame
to the next. On some bias frames, an overall gradient as large as
6 ADU was present across the overscan region, but the trend varied
considerably from one frame to the next, and a fifth order polynomial
was used to fit the overscan.  Fortunately, with only one exception,
we found that the overscan-subtracted bias frame is stable, thus a
useful overscan-subtracted master bias frame could indeed be constructed
from a series of bias exposures, and that master bias is well behaved.
The sole exception is the 25 bias frames from the first night.  Hence we
used the master bias from the second night for the first night as well.

\subsubsection{Variable Slit Width in Final Night} \label{sec:slit}

As mentioned in \S\ref{sec:setup}, a glitch in the slit motor occurred
approximately midway through the last night of observations.  The motor
failed while narrowing the slit back to 225 $\micron$ after we had widened
it to observe a flux standard.  After the problem was fixed and the system
rebooted, we encountered a zeropoint problem in the slit width; the slit
looked visibly wider on the TV monitor for the guider camera. For the
next observation, of the globular cluster NGC 6362, the linewidths in
the arc spectrum are nearly twice as large as in previous observations.
As a result, we narrowed the slit considerably, but ended up with a
narrower slit for the rest of that final night than for the first three
nights. Indeed the lines in the arc spectra are narrower by $\sim$0.2
pixels in FWHM than for the earlier part of the night.  The clusters
affected are NGC 6528 (also observed on a previous night), NGC 6553, NGC
6569, NGC 6752, and 47 Tuc.  Users of the database should be aware that we
have slightly smoothed the spectra of these 5 clusters to give them the
same spectral resolution as the rest of the observations.  The spectrum
of NGC 6362 should only be used for low resolution applications.

\subsubsection{Arc Lamp Issues} \label{sec:arcs}

During the first two nights of the observing run, the rotator mirror
that switches between the main field viewer and the HeNeAr arc lamp was
unreliable, and required frequent rebooting and reinitialization of
the system.  After reinitialization, the wavelength zeropoint could
shift by several pixels, which is not a problem since an arc
lamp exposure was taken along with the object exposure.  However, we
caution the reader that we do not consider the radial velocity data
extracted from our integrated spectra to be reliable.  To circumvent
the rotator mirror problem, for the third and fourth nights the penray
HeNeAr lamp from the Hydra multi-fiber spectrograph was used as the
wavelength calibration source.  This lamp does not provide as uniform
and extensive coverage as the HeNeAr source for the R-C spectrograph,
particularly at the bluer wavelengths.  See \S\ref{sec:reductions}
for a description of how we handled this problem.

\subsubsection{Non-Photometric Conditions} \label{sec:weather}

Although the weather was mostly good during the four-night observing run,
non-photometric conditions were experienced at times during the first and third
nights, i.e., 2003 April 24-25 and 2003 April 26-27.  During the first night,
thin clouds developed during the second half of the night, and appeared to
persist until dawn.  Thus the exposures of NGC~6121, NGC~6266, NGC~6284, 
NGC~6342, and NGC~6626 are all marginally affected by thin clouds on that
night.  On the third night, however, the evening started with thick cirrus,
which eventually developed to the point where we were required to shut down for
approximately an hour.  Thereafter, the sky cleared rapidly, and the conditions
were good for the second half of the night.  Exposures of five clusters,
NGC~1904, NGC~2298, NGC~2808, NGC~5286, and NGC~5904 were taken under quite
heavy clouds, with those of NGC~1904, NGC~2808, and NGC~5286 taken under
approximately a magnitude of extra extinction.  Note that we have at least
one additional exposure for each of these five clusters on other nights.

\section{Data Reduction}\label{sec:reductions}

Analysis of the spectra was carried out using the IRAF software package.
We generally followed the procedures described in the tutorials ``A User's
Guide to CCD Reductions with {\em IRAF}'' by P. Massey and ``A User's
Guide to Reducing Slit Spectra with {\em IRAF}'' by P. Massey, F. Valdes,
and J. Barnes, both available at the IRAF website.  The first step
involved fitting and removing the overscan region and then subtracting
off a master bias frame.  An overscan-subtracted master bias frame was
created for each night from a series of 25 bias frames (taken at the
start of the night).  The 25 frames were averaged, with a {\it minmax}
rejection of the highest and lowest values, using the {\em zerocombine}
routine in the {\em ccdred} package.  The bias-subtracted frames were
then flat-field corrected.  A master flat was produced from a sequence of
25 overscan and bias-subtracted dome flat field frames.  The individual
flats were averaged, and cosmic ray rejection was achieved using the {\em
crreject} parameter in the {\it flatcombine} task. Then the instrumental
and flat field lamp response was normalized along the dispersion direction
using the {\em response} task and the non-uniform illumination along
the slit was measured using a twilight sky exposure according to the
{\em illumination} task.  All object and arc frames were then overscan
and bias subtracted, flat-field divided, and illumination corrected.

Next we extracted a one-dimensional spectrum from the two-dimensional
data frames.  This was accomplished using the {\em apall} task in
the {\em apextract} package.  The extraction aperture was selected by
examining the slit profile for the cluster, and generally included the
core diameter of the cluster.  Two important aspects of the extraction are
to trace the location of the cluster spectrum as a function of position
along the dispersion axis (the {\em aptrace} routine) and eliminating
cosmic rays (CRs) by making a variance-weighted, cleaned extraction
that fits the aperture profile at each wavelength (see the {\em IRAF}
slit spectrum tutorial mentioned above for details).  For a few clusters
(e.g., NGC 2298), with particularly large core radii and/or a jagged
profile along the slit produced by discrete bright stars, it proved
difficult to successfully trace the slit profile at all wavelengths.
In these cases we simply used the trace from a bright star taken on
the same night, but with the aperture resized and initially centered
according to the cluster itself.

The {\em apall} procedure produces not only a variance-weighted, cleaned
extracted spectrum, but an unweighted, uncleaned extracted spectrum, and
a variance spectrum as well.  The variance spectrum was calculated with the
detector read noise and gain properly accounted for.
The variance spectrum for the extracted cluster
spectrum was added in quadrature to the variance spectrum for the extracted
sky spectrum to produce a total variance for the final sky-subtracted cluster
spectrum.  The sky-subtracted cluster spectrum was then divided by the 
variance spectrum to produce a spectrum that gives the S/N ratio at each
pixel.

Wavelength calibration was carried out using an arc spectrum taken at
the same pointing position as the cluster spectrum, either immediately
before or after the cluster and sky background spectra were acquired.
At the start of each night a series of 5 arc spectra were taken, and
averaged to form a master arc spectrum for that night.  The S/N ratio
of the master arc is high enough that many faint lines can be fit, thus
stabilizing the wavelength fitting, especially at the blue end of the
spectrum.  A fourth order Legendre polynomial was fit to the master arc,
generally producing rms residuals of $\sim$0.12 \AA.  Then an extraction
was made from the arc taken along with a particular cluster spectrum,
and the extraction was made with the same trace and aperture parameters
as for the cluster.  Since the S/N ratio of the single arc spectrum is
considerably lower than for the master arc, we then calculated a zero
point shift between the cluster arc and the master arc, using the {\em
reidentify} task.  The assumption is that variations in the geometric
distortion of the spectrograph will be minimal during the night, with only
a zero point shift needing consideration.  The wavelength zero point did
shift significantly during the night, by up to several pixels.  However,
we have verified that the zero point has been succesfully monitored,
since the wavelength of the [\ion{O}{1}]$\lambda$5577 night sky line
in the dispersion corrected spectra is very stable from one cluster
to another.  For the last two nights, when the penray lamp from the
Hydra spectrograph was utilized (cf., \S\ref{sec:arcs}), we used the
master arc from the second night as our high S/N reference arc spectrum.

After the dispersion correction was carried out, the final step was to
flux calibrate the data using the typically 3 flux standards observed
during each night.  Given the low number of flux standards observed,
along with the fact that the mean extinction data for CTIO were used,
and that non-photometric conditions were encountered on some nights
(cf., \S\ref{sec:weather}), we consider the flux calibration to be
only approximate.  The most useful attribute of our spectral library
is the relatively high resolution and high S/N, allowing for numerous
line indices, which do not depend on an absolute flux calibration, to
be defined and used.  Note that our observations do not suffer from the
usual problem encountered in narrow-slit high-resolution spectroscopy,
namely differential atmospheric refraction.  Due to the highly extended
nature of our clusters, coupled with our trailing over the cluster
core, atmospheric dispersion is not a problem.  For our flux standard
observations, we opened the slit to 10\arcs width, to ensure that
atmospheric dispersion did not harm the flux standards either.

We determined radial velocities from cross-correlation of the library
spectra with an integrated spectrum of 47 Tuc obtained in a previous
observing run (see Schiavon et al. 2002a for a description), using the
{\it rv.fxcor} IRAF task. We estimated an internal accuracy of $\pm$9 $km
s^{-1}$ for our radial velocity determinations by measuring the central
wavelengths of absorption lines in the blue and red ends of the spectra.
Finally, these radial velocities were used to bring the 1-D cluster
spectra to the restframe.

\section{Integrated Spectra of Galactic Globular Clusters}\label{sec:spectra}

\subsection{Format of the Spectral Library}\label{sec:format} 

The finalized library of integrated spectra and electronic versions of
Tables 1 and 2
are available at {\tt
http://www.noao.edu/ggclib/}. For each
cluster the final flux-calibrated spectrum is presented in a single
FITS file. Another FITS file, in multi-spectrum format, contains
auxiliary spectral information which may be useful to some investigators.
The multi-spectrum file contains four spectral bands, each one containing
wavelength-calibrated, but {\it not} flux-calibrated, data.  The first
band contains the variance-weighted CR-cleaned, sky-subtracted spectrum,
while the second band contains the unweighted, uncleaned extraction,
both produced by the IRAF {\em apall} task.  The third band contains
the {\em apall} extraction of the sky from the separate sky frame,
except for the few instances where we extracted the sky from the ends
of the slit.  Those exceptions are mentioned in the notes on individual
clusters in \S 6.  The fourth band contains the S/N ratio at each
pixel in the spectrum.  We call the reader's attention for this slight
deviation of the IRAF standard, according to which the fourth band
of {\em apall} output files usually contains the variance spectrum,
instead of the S/N ratio. We decided to adopt the latter in order to
provide the users with a quick visual estimate of the S/N ratio of the
spectra at each pixel.  As mentioned earlier, the S/N ratio per pixel
was obtained by dividing the sky-subtracted cluster spectrum by the
quadrature sum of the variance spectra in the cluster and in the sky.
Therefore, the variance spectrum can be recovered by taking the ratio
between bands 1 and 4 of the auxiliary files.  The README file at the
website gives further details on the format of the spectral data.

\subsection{Examples of the Globular Cluster Spectra} \label{sec:examples}

\begin{figure}
\plotone{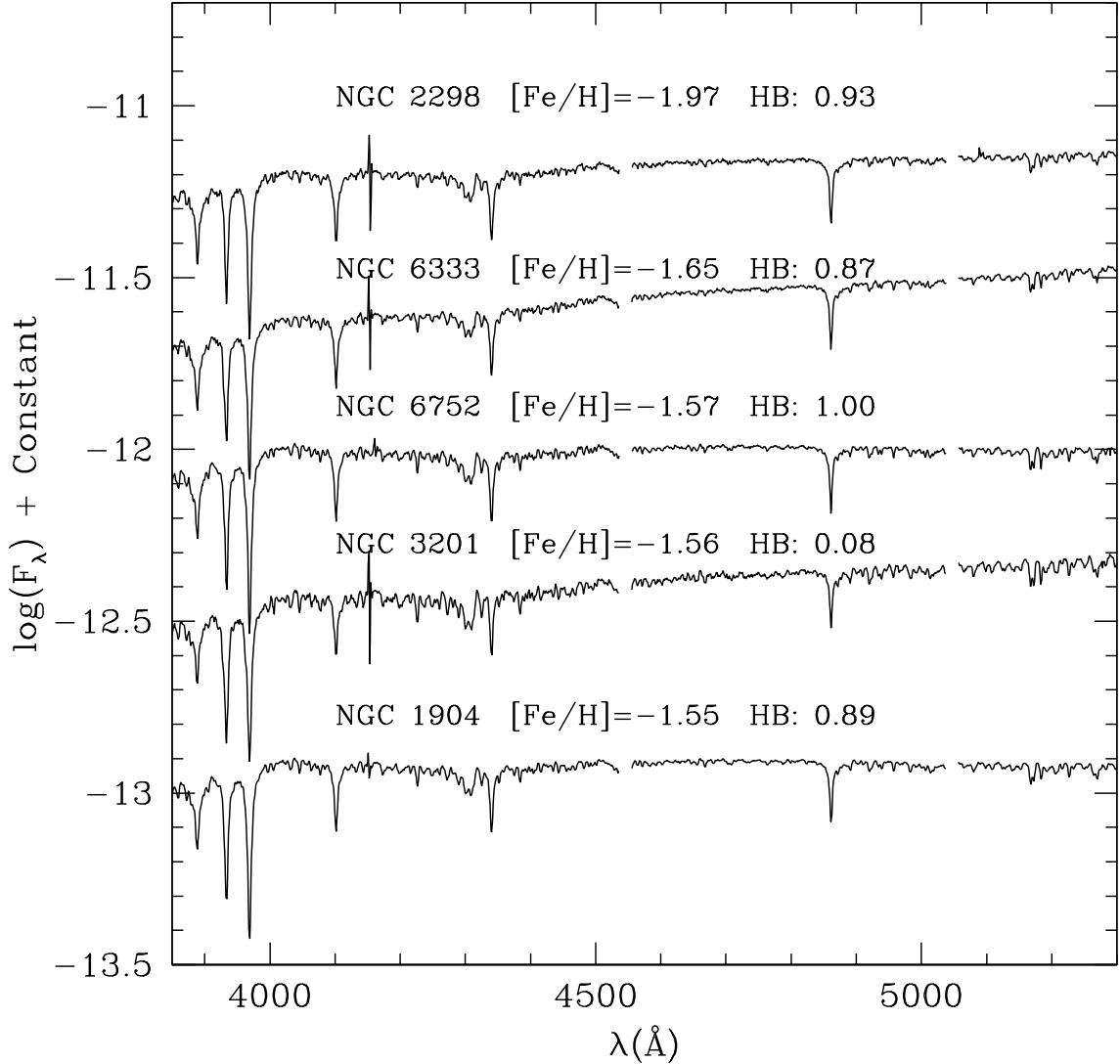}
\caption{a. A segment of the final reduced integrated cluster spectra,
in order of increasing metallicity. For clarity, pixels with very deviant
fluxes due either to CCD defects or to poorly subtracted sky emission
lines were blanked out in this Figure, but preserved in the publicly
available electronic library spectra. The spike present in most spectra
at $\sim$ 4150 ${\rm\AA}$ is due to a CCD defect.}
\label{fig:spectra1}
\end{figure}

Figures~\ref{fig:spectra1}a-h display our complete library of integrated
spectra of Galactic globular clusters, ordered by increasing [Fe/H]. While
the spectral library is clearly intended for work with quantitative
spectral indices, we provide here a few qualitative illustrations of the
use of these spectra. We begin by showing the effect of HB morphology on
the integrated spectra of old stellar populations.  In Figure~\ref{fig:hbm}
we compare the integrated spectra of the two moderately metal-poor GCs
NGC 6723 and NGC 6652. While the two clusters have virtually the same
[Fe/H] (Table~1)
NGC 6652 is characterized by a 
red HB morphology, whereas NGC 6723 has a comparable number of blue and
red HB stars (Piotto \etal 2002). The impact of the blue HB stars in the
integrated spectrum of NGC 6723 can be easily seen by the substantially
{\it stronger} Balmer lines and {\it weaker} metallic features (see also
Schiavon et al. 2004b). Figure~\ref{fig:hbp} shows a similar comparison in
a more metal-rich regime. NGC 6441 and NGC 6304 are more metal-rich than
the above two clusters by $\sim$ 0.5 dex, which is readily noticeable due
to their much stronger metal lines. In NGC 6304, the horizontal branch is
made solely of red stars, while in the case of NGC 6441 it is dominated
by red stars but contains a sizeable blue extension.  The effect of HB
stars on the integrated spectrum of NGC 6441 is more subtle than for
the metal poor clusters discussed above. While Balmer lines are clearly
stronger in the spectrum of NGC~6441, metal lines appear to be only
weakly sensitive to the influence of the blue HB stars, their strengths
being only slightly weaker in the spectrum of NGC~6441, probably because
the horizontal branch of the latter has a dominant red component.


It is well established that the best age indicators for stellar
populations use the sensitivity of Balmer lines to the main sequence
turnoff temperature (Worthey 1994). Hence one must track any conditions
under which anomalous emission lines could lead to an artificial
fill-in of the Balmer lines in the integrated spectrum of a cluster.
Specifically, clusters at low galactic latitude can have a foreground
weak diffuse \ion{H}{2} emission region superimposed on the cluster.
An example of this effect is seen in Fig.~\ref{fig:N3201}, where
the two-dimensional spectrum of NGC~3201 is shown, covering the
[\ion{O}{2}]$\lambda$3727 emission line which extends across the
entire slit. The presence of [\ion{O}{2}] emission implies that
there is emission in the Balmer lines as well. The latter, if not
appropriately subtracted, would fill-in the Balmer lines in absorption,
making them appear artificially weak and thereby making the cluster
appear spuriously old. The identification and removal of any diffuse
emission is thus crucial.  To further emphasize this point with a second
cluster, NGC~6352, which has foreground [\ion{O}{2}] emission, we plot
in Fig.~\ref{fig:n6352spec} the background-subtracted spectrum of NGC~6352
(bottom) and the ``raw'' spectrum before sky subtraction (top). The
[\ion{O}{2}] 3727 emission is readily seen in the raw spectrum, and is
greatly reduced in the subtracted spectrum. (Note that the [\ion{O}{2}]
3727 emission is superposed on a pseudo-continuum peak in the line-rich
absorption spectrum of NGC~6352, making it difficult to assess the
degree of emission contamination without subtracting out the underlying
absorption spectrum of the cluster).

For one last example, we take a closer look at the integrated spectrum
of the relatively metal-rich globular cluster NGC 6637 (= M69). This
cluster contains an emission-line star near the center, as can be seen
in Fig.~\ref{fig:N6637im}, where we show the two-dimensional spectrum
of NGC 6637 in the wavelength region covering H$\delta$ and H$\gamma$.
Aside from a few random CRs, the localized emission in H$\gamma$ and
H$\delta$ at one specific location on the slit (where the slit trailed
across the emission-line star) is clearly evident.  This fact has been
previously noted by Rose \& Tripicco (1986), who tentatively classified
the star as a long-period variable (LPV). In order to further investigate
the nature of the emission-line star we extracted its individual spectrum
and verified that it contains TiO bands whose strength are consistent
with an M2-3 spectral type, so that if the star is a cluster member its
classification as a LPV is reliable. Moreover, the strengths of the Balmer
lines are consistent with those of emission-line M giants found in the
field (e.g. Pereira, Franco \& de Ara\'ujo 2003).  However, we tried to
match the position of the emission-line star in the 2D spectrum with the
positions of known LPV stars in the cluster (Hartwick \& Sandage 1968,
Rosino 1962), without success. We also speculated whether the emission
line star could perhaps be a symbiotic star or a foregroud flaring M
dwarf. However, the spectra of symbiotic stars are characterized by the
presence of strong HeI and [OIII] lines which are absent in our spectrum
(e.g.  Luna \& Costa 2005, Warner 1995). The same applies to the case
of flare M dwarfs, whose spectra are characterized by the presence of
the CaII H and K lines in emission (e.g. Petit 1987), which are also not
seen in the spectrum of the emission-line star in NGC 6637. Therefore,
we suggest that this emission-line star is {\it likely} to be a LPV
member of the cluster.

Given the high luminosity of stars at the giant branch tip, the
strong Balmer emission lines due only to this star can significantly
affect the integrated spectrum of the cluster.  This is shown in
Figure~\ref{fig:N6637spec}, where the variance-weighted ``optimum''
extraction of the integrated spectrum and the unweighted extraction are
both plotted.  Note the major discrepancy in the strengths of H$\delta$
and H$\gamma$ between the two extractions.  The difference is due to the
fact that the variance-weighted extraction identifies and removes the
bright emission spots in the two-dimensional spectrum, just as it does for
CRs. While the identification of the emission lines with CRs is incorrect,
they certainly have that appearance in Figure~\ref{fig:N6637im}, and
their removal from the variance-weighted extraction fortunately avoids
the otherwise catastrophic influence of the LPV emission on the integrated
Balmer lines of the cluster.

The example above raises two important issues. Firstly, under special
circumstances, the influence of a single star can clearly affect the
integrated light of the cluster, even when the entire core diameter
is covered.  Thus it is important to check for any particularly bright
object along the slit. Secondly, our exercise identified a case where
the variance-weighted extraction procedure rejected the localized
emission lines from the LPV as CRs, which was fortunate, but we can
imagine other less benign situations.  Essentially, any bright star in
the cluster with a very different spectrum from the rest of the cluster
is a target for removal by the variance-weighted extraction. For most
clusters, this is not an issue, because the light profile along the slit
is fairly smooth. However, for a few of the less concentrated clusters,
the choice of spectral extraction technique can have an impact on the
final results. For this reason, in the data products we have included
both variance-weighted and unweighted extractions in the multi-spectrum
FITS file for each cluster, as was discussed above.  Thus one can directly
search for anomalies created by the particular extraction technique. Note
that the unweighted extraction has {\it not} been cleaned of CRs.

\section{Notes on Individual Clusters}\label{sec:notes}

In this section we provide information on individual clusters.

\noindent {\bf NGC~104 = 47 Tuc:} 47 Tuc is so bright that no separate sky
exposure was taken. This is one of the clusters observed during the
last night when the slit width was slightly narrower than for the other
observations.  We have smoothed the final spectrum of this cluster to match
the spectral resolution of the other clusters.

\noindent {\bf NGC~1904:} It was difficult to obtain a successful aperture 
trace for this cluster, due to the presence of a very red star that traversed
the slit right at the center of the cluster.  However, by adjusting the
step size of the aperture trace, a successful trace was carried out.  

\noindent {\bf NGC~2298:} NGC~2298 could not be successfully traced for 
aperture extraction, due to a large number of discrete bright stars crossing 
the slit, and creating a jagged slit profile.  Consequently, we used the
aperture trace from a stellar spectrum taken on the same night, but recentered 
the aperture and resized it.  There are two spectra of this cluster, taken on
different nights, with very slightly different aperture extractions.

\noindent {\bf NGC~2808:} Two exposures were taken, heavy clouds compromise
the second one.  We used slightly different aperture extractions for the
two exposures.

\noindent {\bf NGC~3201:} NGC~3201 is a cluster with a particularly
large core diameter.  It is somewhat difficult to determine the cluster
center.  We obtained spectra for the cluster on both UT 25-APR-2003 and UT
26-APR-2003.  There is strong [OII]$\lambda$3727 emission evident across
the entire slit in both cluster exposures and the two sky background
regions scanned (the sky background regions scanned were different on
the two nights, as is recorded in Table~2. 
An extremely
bright star crossed the slit during the sky background trail for the
exposure on UT 25-APR-2003.  Consequently, we decreased the extracted
aperture on the sky background frame to $\pm$55 pixels ($\pm55$\arcs),
to avoid the contaminated location on the slit.  We then renormalized
the sky counts to match the $\pm$85 pixel extraction on the cluster.
This problem did not occur for the exposure on UT 26-APR-2003, for
which we scanned a different sky background region.  Note that the
[OII]$\lambda$3727 emission subtraction appears to come out cleaner for
the UT 26-APR-2003 exposure.

\noindent {\bf NGC~5286:} There is no arc for NGC~5286 on the night of UT 
2003-04-27, so we used the arc for NGC~2808, and made a zeropoint correction 
using night sky lines.

\noindent {\bf NGC~5904:} NGC~5904 could not be successfully traced at all, 
maybe a result of too many discrete stars crossing the slit
  with different SED's.  So we used the trace from the star
  HD102574, but recentered the aperture and resized it.

\noindent {\bf NGC~5927:} NGC~5927 was difficult to trace, so used trace from 
the star HD76151, but recentered the aperture and resized it.

\noindent {\bf NGC~5986:} The trail across NGC~5986 intersected a very bright
star.  We did not choose to eliminate the bright star from our aperture
extraction.

\noindent {\bf NGC~6171:} There is extended [OII]$\lambda$3727 emission across
this spatially extended cluster. To obtain a successful aperture trace for the 
cluster it was necessary to center the extracted aperture on the 
brightest peak on the slit profile rather than on the average center of the
light distribution.

\noindent {\bf NGC~6284:} NGC~6284 is so centrally concentrated that we defined 
two apertures.  The first aperture contains the FWHM of the slit profile, while
the second contains a substantially wider region along the slit.

\noindent {\bf NGC~6342:} NGC~6342 is so centrally concentrated that we defined
two apertures.  The first aperture contains the FWHM of the slit profile, while
the second contains a substantially wider region along the slit. 

\noindent {\bf NGC~6544:} NGC~6544 has an extremely peaked light profile along
the slit that is offset by several arcseconds from the centroid of the
fainter light levels.  We used the peak in the light profile for the center
of the extracted aperture, but used asymmetric limits for the aperture.

\noindent {\bf NGC~6352:}  NGC~6352 has emission at [OII]$\lambda$3727.
Sky subtraction in [OII] emission came out better when the background was
subtracted from the ends of the slit rather than from the separate sky
background exposure.  The background in the separate frame was further
compromised by several bright stars that went through the slit in the
5\arcm trail. Hence we used the ends of the slit for sky subtraction in
this case.

\noindent {\bf NGC~6362:}  The spectrun for this cluster has degraded 
resolution, due to the problem of the slit width control described in
\S\ref{sec:slit}.

\noindent {\bf NGC~6441:} NGC~6441 is so centrally concentrated that we defined
two apertures.  The first aperture contains the FWHM of the slit profile, while
the second contains a substantially wider region along the slit. 

\noindent {\bf NGC~6528:} For the 28-APR-2004 observations of NGC~6528
inaccuracy in the tracking led to the 15 minute trail not
covering the anticipated 9\arcsc.  As a result we took a second 15 minute
exposure, which picked up the trail where it ended in the first exposure.  The
counts in the second exposure are very similar to those int he first
exposure.  We have maintained these as two separate observations
(NGC6528b*.fits and NGC6528c*.fits) because they
are taken from different parts of the cluster.  In addition, NGC~6528 is so 
centrally concentrated that we defined
two apertures.  The first aperture contains the FWHM of the slit profile, while
the second contains a substantially wider region along the slit.  Finally.
we have smoothed the final spectra of this cluster taken on 28-APR-2004 to
match the spectral resolution of the other clusters.

\noindent {\bf NGC~6553:} This is perhaps the most problematic cluster in the
sample.  It is in a high background region of the bulge, with variable
obscuration and variable [OII] emission.  In addition, NGC~6553 is not a
centrally concentrated cluster.  Consequently, it was a difficult cluster to
trace, so we used the trace from a star, but with redefined apertures.  
We found that the [OII] and Balmer emission in the background is stronger
in the separate sky exposure than at the ends of the slit for the cluster
exposure.  Consequently, we used the ends of the slit for background
subtraction.  In addition, there is a bright blue foreground star that
traversed the slit during the cluster exposure.  This star is bright enough
to make an appreciable impact in strengthening the Balmer lines in the
integrated cluster spectrum.  Hence we separately extracted the slit position
corrsponding tot he bright star and subtracted that off.  Nevertheless, even
when the bright star contributioin is removed, the relative depths of Ca II
H and K lines indicates a contribution from hot stars to the integrated
spectrum of this cluster.  We believe this reflects the contribution of blue
straggler stars that are seen in the color-magnitude diagram of the
cluster.

\noindent {\bf NGC~6569:} NGC~6569 is one of the clusters observed during the
last night when the slit width was slightly narrower than for the other
observations.  We have smoothed the final spectrum of this cluster to match
the spectral resolution of the other clusters.

\noindent {\bf NGC~6624:} NGC~6624 is so centrally concentrated that we defined
two apertures.  The first aperture contains the FWHM of the slit profile, while
the second contains a substantially wider region along the slit.

\noindent {\bf NGC~6637:} NGC~6637 has a long period variable in emission in 
the cluster core.  The cleaned, variance-weighted extraction is unaffected by 
this (the LPV Balmer line emission is confused with cosmic rays and cleaned out), 
but the raw counts extraction has the Balmer lines seriously weakened by 
the emission in the star.  See \S\ref{sec:examples} for further details.

\noindent {\bf NGC~6652:} For this cluster either the discrete digitization of
the trail rate or inaccuracy in the tracking led to the 15 minute trail not
covering the aniticipated 9\arcsc.  As a result we took a second 15 minute
exposure, which picked up the trail where it ended in the first exposure.  The
counts in the second exposure are $\sim$30\% higher than in the first 
exposure.  We have maintained these as two separate observations because they
are taken from different parts of the cluster.

\noindent {\bf NGC~6723:} We took a 10 minute exposure on the cluster and
only a 5 minute sky exposure, thus doubled the sky counts before subtracting
them from the cluster.

\noindent {\bf NGC~6752:} NGC~6752 is one of the clusters observed during the
last night when the slit width was slightly narrower than for the other
observations.  We have smoothed the final spectrum of this cluster to match
the spectral resolution of the other clusters.

\noindent {\bf NGC~7078:} NGC~7078 is so centrally concentrated that we defined
two apertures.  The first aperture contains the FWHM of the slit profile, while
the second contains a substantially wider region along the slit.

\noindent {\bf NGC~7089:} We used sky from the ends of the slit, rather than
from the subsequent sky exposure, because the
sky background was starting to brighten considerably with approaching dawn.

\section{Summary} \label{sec:summary}

We have collected a new library of integrated spectra of Galactic globular
clusters, with high S/N and moderately high resolution. We envisage at
least two chief applications for this new dataset. Firstly, it should
serve as a reliable database for local, well-known stellar populations,
against which the observations of remote, unresolved, systems can be
contrasted. Secondly, this library can be used to verify the accuracy
and reliability of the predictions of stellar population synthesis
models. For both purposes, it is crucial that the fundamental parameters
of the clusters that make up the library be well known, and for that
reason we selected our targets amongst those with best color-magnitude
and metal-abundance data.  Further applications aiming at a detailed
calibration of stellar population synthesis models are currently under
way.  The spectral library is publicly available in electronic format
from the National Optical Astronomical Observatory website.

\acknowledgements{We thank the referee, Guy Worthey, for suggestions
that improved this paper. R.P.S. acknowledges financial support from
NSF grant AST 00-71198 to the University of California-Santa Cruz,
and from HST Treasury Program grant GO-09455.05-A to the University of
Virginia. S.C. and L.A.M. are grateful to the NSERC for financial support
through a Discovery grant. This research was partially supported by NSF
grant AST-0406443 to the University of North Carolina-Chapel Hill.}

\newpage

\clearpage

\begin{deluxetable}{llrrcrrcccl}  
\label{tab:mastertable}
\tabletypesize{\scriptsize}
\tablecaption{Sample Clusters}
\tablewidth{0pt}
\tablehead{
\colhead{NGC}     & \colhead{Other} & \colhead{$l$}   & \colhead{$b$} & 
\colhead{$R_{GC}$}& \colhead{$r_c$}  & \colhead{c}   & \colhead{$E(B-V)$} &
\colhead{[Fe/H]} & \colhead{${(B-R)\over (B+V+R)}$}     & \colhead{CMD Ref.} \\
\colhead{(1)}     & \colhead{(2)} & \colhead{(3)}   & \colhead{(4)} & 
\colhead{(5)}& \colhead{(6)}  & \colhead{(7)}   & \colhead{(8)} &
\colhead{(9)} & \colhead{(10)}     & \colhead{(11)} \\
}
\startdata
104     &47 Tuc&305.90 &-44.89 & 7.4 & 0.40 & 2.03 & 0.04 & -0.70$^b$ & -0.99 & 1,2,12 \\
1851    &      &244.51 &-35.04 &16.7 & 0.06 & 2.32 & 0.02 & -1.21$^d$ & -0.36 & 1,2 \\
1904    & M 79 &227.23 &-29.35 &18.8 & 0.16 & 1.72 & 0.01 & -1.55$^d$ & -0.89 & 1,2 \\
2298$^a$&      &245.63 &-16.01 &15.1 & 0.34 & 1.28 & 0.15 & -1.97$^b$ &  0.93 & 2 \\
2808    &      &282.19 &-11.25 &11.0 & 0.26 & 1.77 & 0.23 & -1.29$^d$ & -0.49 & 1,2 \\
3201    &      &277.23 &  8.64 & 9.0 & 1.43 & 1.30 & 0.21 & -1.56$^b$ &  0.08 & 1,2,14 \\
5286$^a$&      &311.61 & 10.57 & 7.2 & 0.29 & 1.46 & 0.24 & -1.51$^e$ &  0.80 & 4 \\
5904    & M 5  &  3.86 & 46.80 & 6.2 & 0.42 & 1.83 & 0.03 & -1.26$^b$ &  0.31 & 1,3,13,21 \\
5927    &      &326.60 &  4.86 & 4.5 & 0.42 & 1.60 & 0.45 & -0.64$^c$ & -1.00 & 1,2 \\
5946    &      &327.58 &  4.19 & 7.4 & 0.08 & 2.50 & 0.54 & -1.54$^e$ &  B$^f$ & 1 \\
5986    &      &337.02 & 13.27 & 4.8 & 0.63 & 1.22 & 0.27 & -1.53$^d$ &  0.97 & 1,2,17 \\
6121$^a$& M 4  &350.97 & 15.97 & 6.2 & 0.83 & 1.59 & 0.40 & -1.15$^b$ & -0.06 & 2,5 \\
6171    & M 107&  3.37 & 23.01 & 3.3 & 0.54 & 1.51 & 0.33 & -1.13$^d$ & -0.73 & 1,2 \\
6218    & M 12 & 15.72 & 26.31 & 4.5 & 0.72 & 1.39 & 0.19 & -1.32$^d$ &  0.97 & 1,3,15,16 \\
6235    &      &358.92 & 13.52 & 2.9 & 0.13 & 1.33 & 0.36 & -1.36$^d$ &  0.89 & 1 \\
6254$^a$& M 10 & 15.14 & 23.08 & 4.7 & 0.86 & 1.40 & 0.28 & -1.51$^b$ &  0.98 & 3,15 \\
6266    & M 62 &353.58 &  7.32 & 1.7 & 0.18 & 1.70 & 0.47 & -1.20$^d$ &  0.32 & 1,2 \\
6284    &      &358.35 &  9.94 & 6.9 & 0.07 & 2.50 & 0.28 & -1.27$^e$ &  B$^f$  & 1 \\
6304    &      &355.83 &  5.38 & 2.1 & 0.21 & 1.80 & 0.52 & -0.66$^c$ & -1.00 & 1,2 \\
6316    &      &357.18 &  5.76 & 3.2 & 0.17 & 1.55 & 0.51 & -0.90$^e$ & -1.00 & 1 \\
6333$^a$& M 9  &  5.54 & 10.71 & 1.7 & 0.58 & 1.15 & 0.35 & -1.65$^e$ &  0.87 & 6 \\
6342    &      &  4.90 &  9.73 & 1.7 & 0.05 & 2.50 & 0.46 & -1.01$^e$ & -1.00 & 1 \\
6352$^a$&      &341.42 & -7.17 & 3.1 & 0.83 & 1.10 & 0.19 & -0.70$^c$ & -1.00 & 2,7,8 \\
6356    &      &  6.72 & 10.22 & 7.6 & 0.23 & 1.54 & 0.28 & -0.74$^e$ & -1.00 & 1 \\
6362    &      &325.55 &-17.57 & 5.3 & 1.32 & 1.10 & 0.08 & -1.17$^d$ & -0.58 & 1,2,18 \\
6388    &      &345.56 & -6.74 & 4.4 & 0.12 & 1.70 & 0.40 & -0.68$^e$ & I$^f$  & 1 \\
6441    &      &353.53 & -5.01 & 3.5 & 0.11 & 1.85 & 0.44 & -0.65$^e$ & I$^f$  & 1 \\
6522    &      &  1.02 & -3.93 & 0.6 & 0.05 & 2.50 & 0.48 & -1.39$^d$ &  0.71 & 1 \\
6528$^a$&      &  1.14 & -4.17 & 1.5 & 0.09 & 2.29 & 0.62 & -0.10$^d$ & -1.00 & 9,19 \\
6544    &      &  5.84 & -2.20 & 5.4 & 0.05 & 1.63 & 0.73 & -1.38$^d$ &  1.00 & 1,2 \\
6553$^a$&      &  5.25 & -3.02 & 4.6 & 0.55 & 1.17 & 0.80 & -0.20$^d$ & -1.00 & 10 \\
6569    &      &  0.48 & -6.68 & 1.2 & 0.37 & 1.27 & 0.56 & -1.08$^e$ & MR$^f$  & 1,20 \\
6624    &      &  2.79 & -7.91 & 1.2 & 0.06 & 2.50 & 0.28 & -0.70$^c$ & -1.00 & 1 \\
6626$^a$&M 28  &  7.80 & -5.58 & 2.4 & 0.24 & 1.67 & 0.38 & -1.21$^d$ &  0.90 & 2 \\
6637    &M 69  &  1.72 &-10.27 & 1.6 & 0.34 & 1.39 & 0.16 & -0.78$^c$ & -1.00 & 1,2 \\
6638    &      &  7.90 & -7.15 & 1.6 & 0.26 & 1.40 & 0.40 & -1.08$^d$ & -0.30 & 1,2 \\
6652    &      &  1.53 &-11.38 & 2.4 & 0.07 & 1.80 & 0.09 & -1.10$^e$ & -1.00 & 1 \\
6723    &      &  0.07 &-17.30 & 2.6 & 0.94 & 1.05 & 0.05 & -1.14$^d$ & -0.08 & 1,2 \\
6752$^a$&      &336.50 &-25.63 & 5.1 & 0.17 & 2.50 & 0.04 & -1.57$^b$ &  1.00 & 2,11 \\
7089    &M 2   & 53.38 &-35.78 &10.4 & 0.34 & 1.80 & 0.06 & -1.49$^e$ &  0.96 & 1 \\
\enddata
\tablenotetext{a}{Galactic coordinates, Galactocentric distances and
reddenings come from Djorgovski \& Meylan (1993), Djorgovski (1993),
and Peterson (1993)}
\tablenotetext{b}{[Fe/H] from Kraft \& Ivans (2003); $^c$[Fe/H] from Carretta \& 
Gratton (1997); $^d$[Fe/H] from Carretta \& Gratton (1997); corrected by -0.18 
dex (see text); $^e$[Fe/H] from Schiavon \etal (2005, in preparation)}
\tablenotetext{f}{B: blue HB, no clear red HB stars seen; I: intermediate
type HB, with pronounced red clump plus an extended blue tail; MR:
mostly red, with a sparse blue component}
\tablecomments{References to color-magnitude diagrams in the literature:
1) Piotto \etal (2002); 2) Rosenberg \etal (2000a); 3) Rosenberg \etal
(2000b); 4) Samus \etal (1995); 5) Kanatas \etal (1995); 6) Janes \&
Heasley (1991); 7) Pulone \etal (2003); 8) Fullton \etal (1995); 9)
Feltzing \& Johnson (2002); 10) Zoccali \etal (2001); 11) Momany \etal
(2002); 12) Howell et al. (2000), 13) Markov et al. (2001); 14) von Braun
\& Mateo (2001); 15) von Braun et al. (2002); 16) Hargis et al. (2004);
17) Ortolani et al. (2000); 18) Brocato et al. (1999); 19) Momany \etal
(2003); 20) Ortolani et al. (2001); 21) Sandquist et al. (1996).}

\end{deluxetable}
\clearpage

\begin{deluxetable}{lrrrrrrrrr} \label{tab:journal}
\tabletypesize{\footnotesize}
\tablewidth{0pt}
\tablecolumns{9}

\tablecaption{Journal of Observations for Galactic Globular Clusters}
\tablehead{
\colhead{(1)} &
\colhead{(2)} &
\colhead{(3)} &
\colhead{(4)} &
\colhead{(5)} &
\colhead{(6)} &
\colhead{(7)} &
\colhead{(8)} &
\colhead{(9)}
}

\startdata
NGC~104 & 2003-04-28 & 5 & E-W & 48\arcs & \nodata & \nodata & \nodata & $\pm$25\arcsc \\
NGC~1851 & 2003-04-25 & 5 & N-S & 9\arcs & 5 & 20\arcm W & \nodata & $\pm$9\arcsc \\
NGC~1904 & 2003-04-26 & 15 & E-W & 20\arcs & 15 & 15\arcm E & \nodata & $\pm$10\arcsc \\
NGC~1904 & 2003-04-27 & 15 & E-W & 20\arcs & 15 & 15\arcm E & \nodata & $\pm$14\arcsc \\
NGC~2298 & 2003-04-27 & 15 & E-W & 36\arcs & 15 & 15\arcm E & \nodata & -24\arcsc, +20\arcsc \\
NGC~2298 & 2003-04-28 & 15 & E-W & 36\arcs & 15 & 15\arcm E & \nodata & $\pm$20 \\
NGC~2808 & 2003-04-25 & 10 & E-W & 30\arcs & 5 & 15\arcm E & \nodata & $\pm$20 \\
NGC~2808 & 2003-04-27 & 10 & E-W & 30\arcs & 10 & 15\arcm W & 5\arcm & -17\arcsc, +15\arcsc \\
NGC~3201 & 2003-04-25 & 15 & N-S & 170\arcs & 15 & 12\arcm W & 5\arcm & $\pm$85\arcsc \\
NGC~3201 & 2003-04-26 & 15 & N-S & 170\arcs & 15 & 12\arcm N & 5\arcm & $\pm$85\arcsc \\
NGC~5286 & 2003-04-25 & 10 & N-S & 36\arcs & 10 & 12\arcm N & 5\arcm & $\pm$20\arcsc \\
NGC~5286 & 2003-04-27 & 10 & E-W & 36\arcs & 10 & 12\arcm W & 5\arcm & $\pm$20\arcsc \\
NGC~5286 & 2003-04-28 & 15 & E-W & 36\arcs & 15 & 12\arcm W & 5\arcm & -20\arcsc, +23\arcsc \\
NGC~5904 & 2003-04-27 & 15 & E-W & 54\arcs & 15 & 20\arcm W & 5\arcm & -27\arcsc, +30\arcsc \\
NGC~5904 & 2003-04-28 & 15 & E-W & 45\arcs & 15 & 20\arcm N & 5\arcm & -27\arcsc
, +30\arcsc \\
NGC~5927 & 2003-04-25 & 15 & N-S & 50\arcs & 15 & 20\arcm W & 10\arcm & $\pm$29\\
NGC~5927 & 2003-04-25 & 15 & N-S & 50\arcs & 15 & 15\arcm E & 10\arcm & $\pm$29\\
NGC~5927 & 2003-04-28 & 15 & E-W & 46\arcs & 15 & 15\arcm E & 10\arcm & $\pm$25\\
NGC~5946 & 2003-04-26 & 15 & N-S & 10\arcs & 15 & 10\arcm W & 10\arcm & -6\arcsc, +7\arcsc \\
NGC~5986 & 2003-04-25 & 15 & N-S & 76\arcs & 15 & 12\arcm N & 5\arcm & -32\arcsc, +42\arcsc \\
NGC~6121 & 2003-04-25 & 15 & N-S & 100\arcsc & 15 & 14\arcm S, 14\arcm W & 10\arcm & $\pm$50\arcsc \\
NGC~6171 & 2003-04-26 & 15 & N-S & 64\arcsc & 15 & 11\arcm N, 11\arcm W & 10\arcm & -32\arcsc, +36\arcsc \\
NGC~6171 & 2003-04-26 & 15 & N-S & 64\arcsc & 15 & 10\arcm S & 10\arcm & -32\arcsc, +37\arcsc \\
NGC~6218 & 2003-04-27 & 15 & E-W & 90\arcsc & 15 & 20\arcm W & 10\arcm & $\pm$45\arcsc \\
NGC~6235 & 2003-04-28 & 15 & E-W & 45\arcsc & 15 & 20\arcm N & 10\arcm & $\pm$25\arcsc \\
NGC~6254 & 2003-04-27 & 15 & E-W & 108\arcsc & 15 & 15\arcm N & 10\arcm & $\pm$54\arcs \\
NGC~6266 & 2003-04-25 & 15 & N-S & 22\arcsc & 15 & 15\arcm N & 5\arcm & -12\arcsc, +16\arcsc \\
NGC~6284 & 2003-04-25 & 15 & N-S & 9\arcsc & 15 & 15\arcm N & 5\arcm & -6\arcsc, +7\arcsc \\
\nodata & \nodata & \nodata & \nodata & \nodata & \nodata &\nodata & \nodata & $\pm$14\arcsc \\
NGC~6304 & 2003-04-26 & 15 & N-S & 26\arcsc & 15 & 10\arcm N & 5\arcm & $\pm$17\arcsc \\
NGC~6316 & 2003-04-26 & 15 & N-S & 20\arcsc & 15 & 10\arcm E & 5\arcm & -10\arcsc, +11\arcsc \\
NGC~6316 & 2003-04-28 & 15 & E-W & 18\arcsc & 15 & 10\arcm W & 5\arcm & -14\arcsc, +11\arcsc \\
NGC~6333 & 2003-04-26 & 15 & N-S & 70\arcsc & 15 & 15\arcm E & 5\arcm & $\pm$35\arcsc \\
NGC~6342 & 2003-04-25 & 10 & N-S & 12\arcsc & 10 & 10\arcm S & 5\arcm & $\pm$4\arcsc \\
\nodata & \nodata & \nodata & \nodata & \nodata & \nodata &\nodata & \nodata & $\pm$10\arcsc \\
NGC~6352 & 2003-04-27 & 15 & E-W & 99\arcsc & \nodata & \nodata &\nodata & $\pm$50\arcsc \\
NGC~6356 & 2003-04-26 & 15 & N-S & 28\arcsc & 15 & 15\arcm E & 5\arcm & $\pm$22\arcsc \\
NGC~6362 & 2003-04-28 & 15 & E-W & 160\arcsc & 15 & 15\arcm W & 5\arcm & $\pm$80\arcsc \\
NGC~6388 & 2003-04-26 & 10 & N-S & 18\arcsc & 10 & 10\arcm E & 5\arcm & $\pm$10\arcsc \\
NGC~6441 & 2003-04-26 & 10 & N-S & 12\arcsc & 10 & 10\arcm E & 5\arcm & $\pm$10\arcsc \\
\nodata & \nodata & \nodata & \nodata & \nodata & \nodata &\nodata & \nodata & $\pm$18\arcsc \\
NGC~6522 & 2003-04-26 & 15 & N-S & 9\arcsc & 15 & 8\arcm S & 5\arcm & $\pm$9\arcsc \\
NGC~6528 & 2003-04-27 & 15 & E-W & 9\arcsc & 15 & 8\arcm N & 5\arcm & $\pm$8\arcsc \\
\nodata & \nodata & \nodata & \nodata & \nodata & \nodata &\nodata & \nodata & $\pm$15\arcsc \\
NGC~6528\tablenotemark{a} & 2003-04-28 & 15 & E-W & 5\arcsc & 15 & 15\arcm W & 5\arcm & $\pm$7\arcsc \\
\nodata & \nodata & \nodata & \nodata & \nodata & \nodata &\nodata & \nodata & $
\pm$14\arcsc \\
NGC~6544 & 2003-04-26 & 15 & N-S & 18\arcsc & 15 & 15\arcm E & 10\arcm & -12\arcsc, +6\arcsc \\
NGC~6553 & 2003-04-28 & 15 & E-W & 64\arcsc & 15 & \nodata & \nodata & $\pm$35\arcsc \\
NGC~6569 & 2003-04-28 & 15 & E-W & 45\arcsc & 15 & 8\arcm W, 8\arcm S & 10\arcm & $\pm$25\arcsc \\
NGC~6624 & 2003-04-27 & 15 & E-W & 9\arcsc & 15 & 15\arcm S & 5\arcm & -7\arcsc, +5\arcsc \\
\nodata & \nodata & \nodata & \nodata & \nodata & \nodata &\nodata & \nodata & $\pm$11\arcsc \\
NGC~6626 & 2003-04-25 & 10 & N-S & 28\arcsc & 10 & 15\arcm S & 5\arcm & $\pm$16\arcsc \\
NGC~6637 & 2003-04-27 & 15 & E-W & 36\arcsc & 15 & 15\arcm N & 10\arcm & $\pm$20\arcsc \\
NGC~6638 & 2003-04-27 & 15 & E-W & 27\arcsc & 15 & 15\arcm N & 5\arcm & $\pm$15\arcsc \\
NGC~6652 & 2003-04-27 & 15 & E-W & 9\arcsc & 15 & 12\arcm W & 5\arcm & $\pm$7\arcsc \\
NGC~6723 & 2003-04-27 & 10 & E-W & 108\arcsc & 5 & 12\arcm W & 5\arcm & $\pm$54\arcsc \\
NGC~6752 & 2003-04-28 & 5 & E-W & 18\arcs & 5 & 12\arcm W & \nodata & -13\arcs, +11\arcsc \\
NGC~7078 & 2003-04-26 & 5 & N-S & 9\arcs & \nodata & \nodata & \nodata & $\pm$5\arcsc \\
\nodata & \nodata & \nodata & \nodata & \nodata & \nodata &\nodata & \nodata & $\pm$20\arcsc \\
NGC~7089 & 2003-04-28 & 5 & E-W & 36\arcsc & \nodata & \nodata & \nodata &\nodata $\pm$22\arcsc \\

\enddata
\tablenotetext{a}{As discussed in the Notes on this cluster, there were two
consecutive exposures made for NGC~6528 on this night, both with the same
two aperture sizes.}
\end{deluxetable}

\clearpage

\setcounter{figure}{1}
\begin{figure}
\plotone{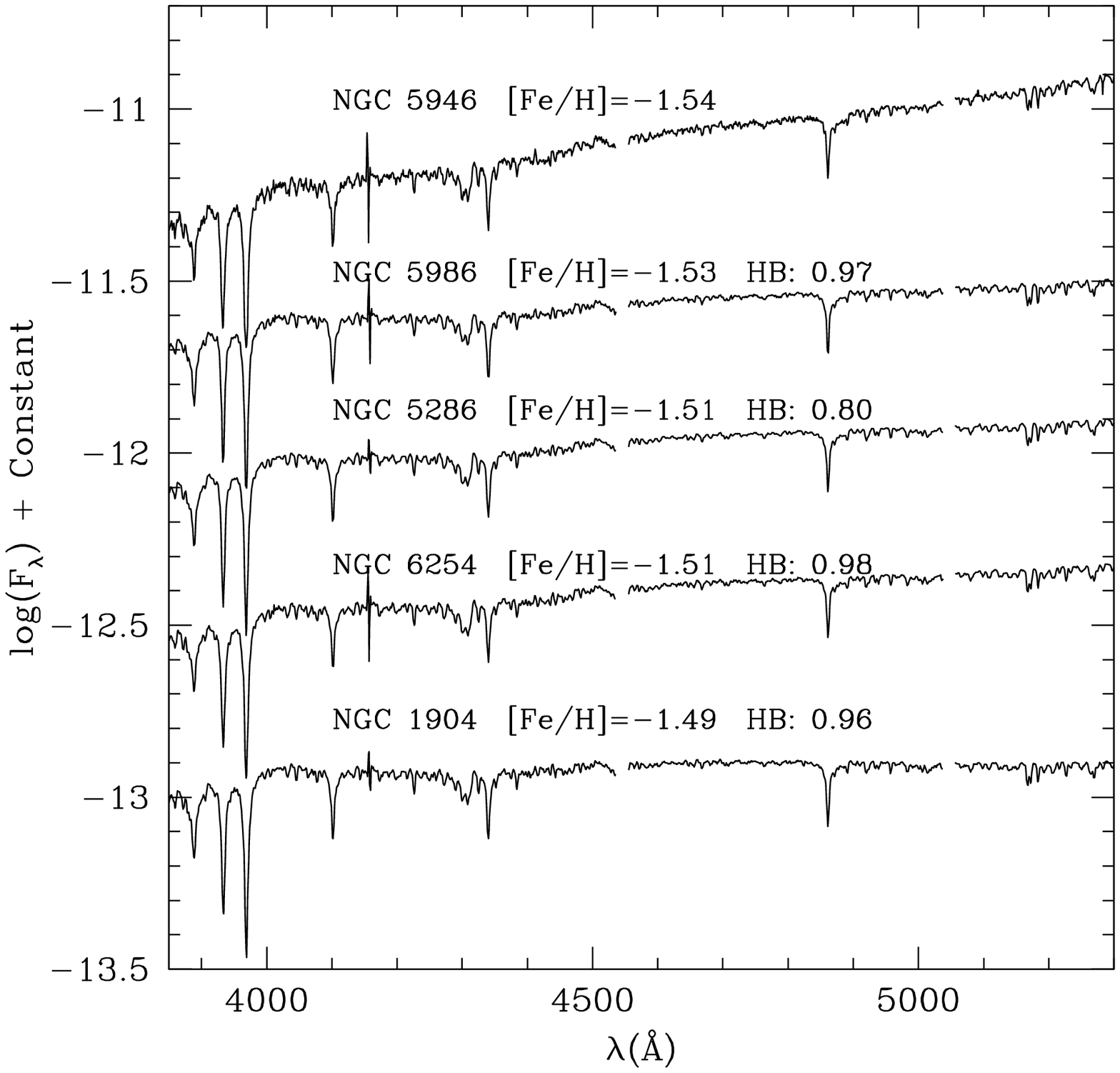}
\caption{b. Cont.}
\label{fig:spectra2}
\end{figure}
\setcounter{figure}{1}
\begin{figure}
\plotone{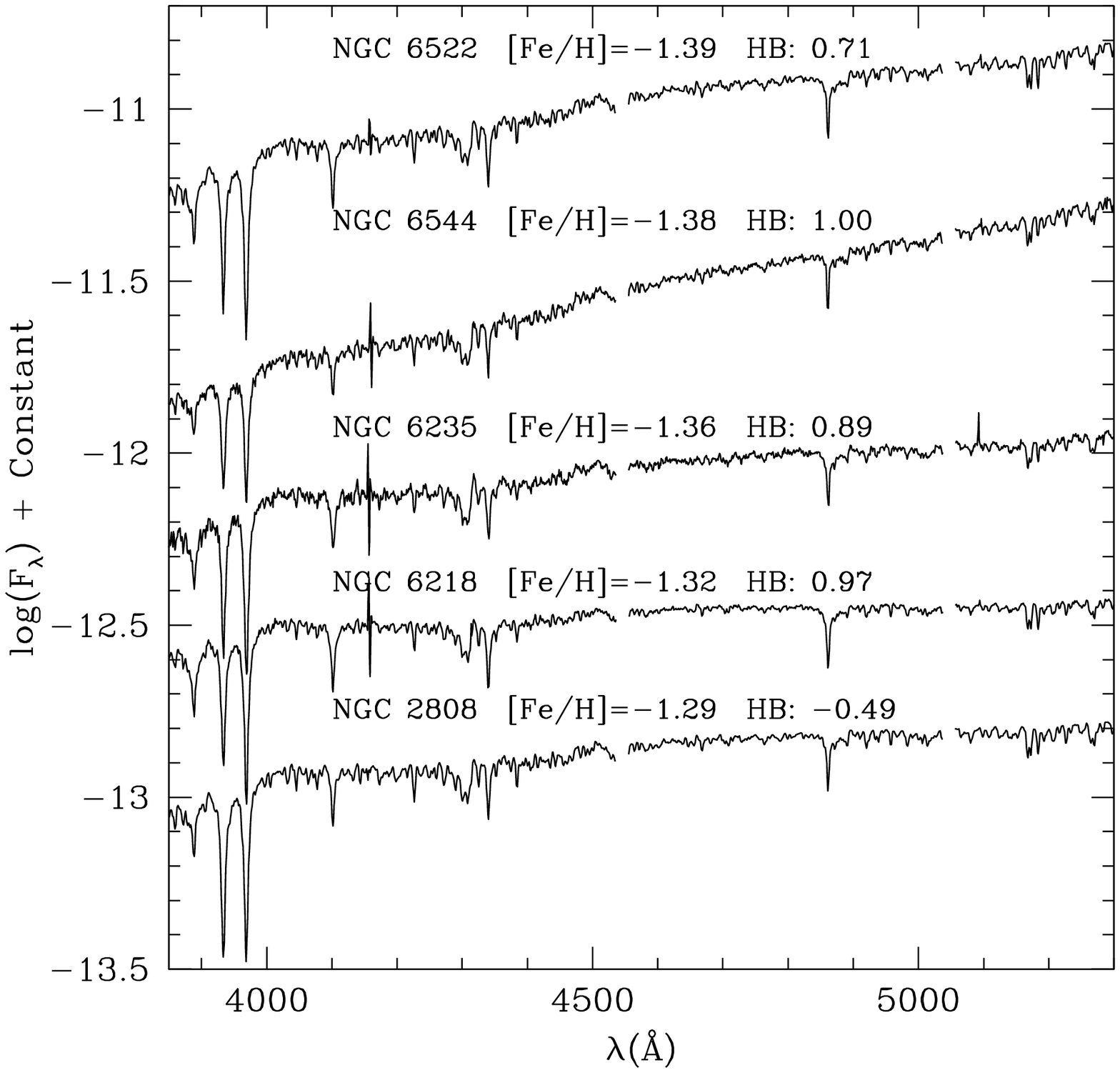}
\caption{c. Cont.}
\label{fig:spectra3}
\end{figure}
\setcounter{figure}{1}
\begin{figure}
\plotone{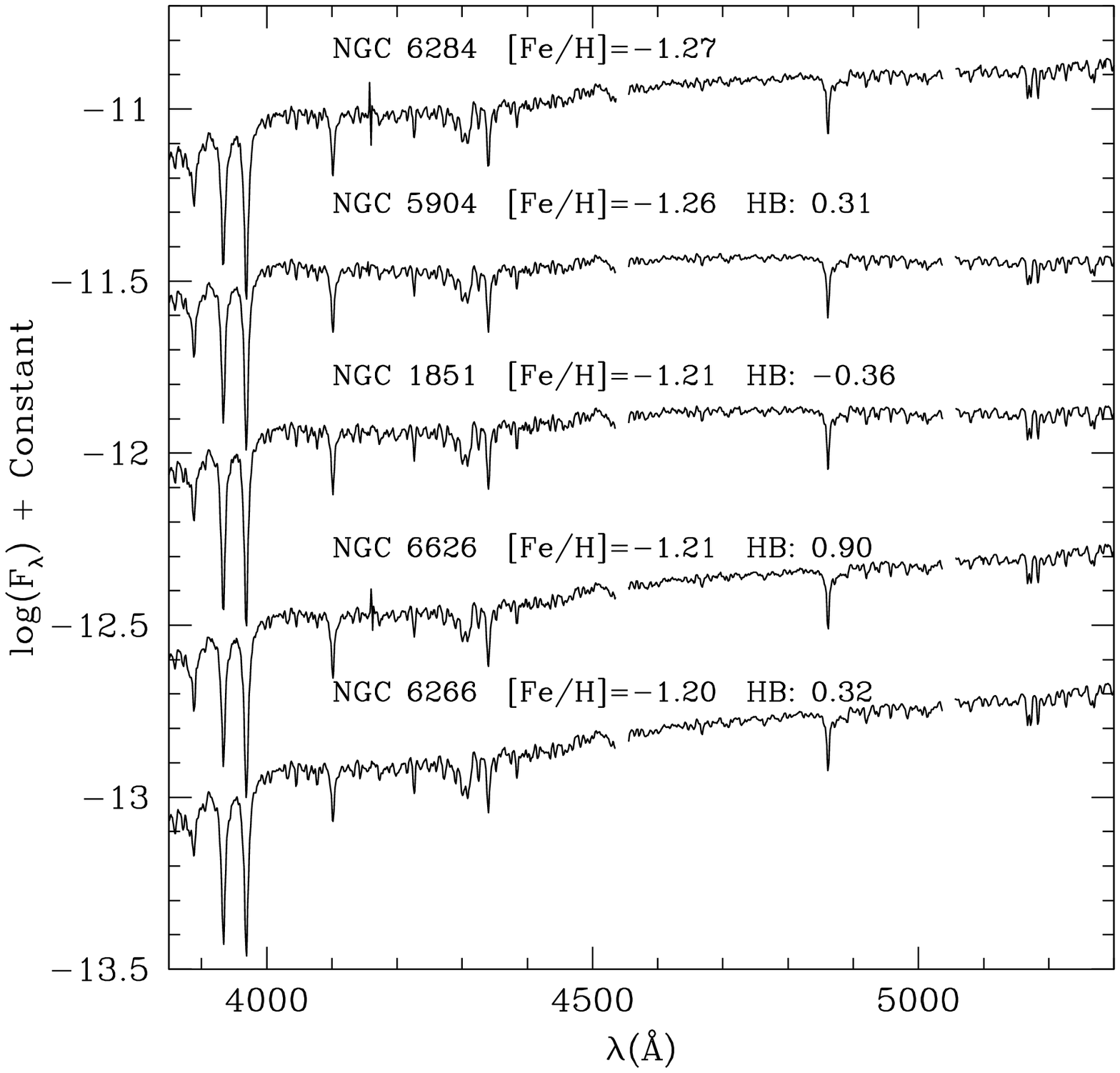}
\caption{d. Cont.}
\label{fig:spectra4}
\end{figure}
\setcounter{figure}{1}
\begin{figure}
\plotone{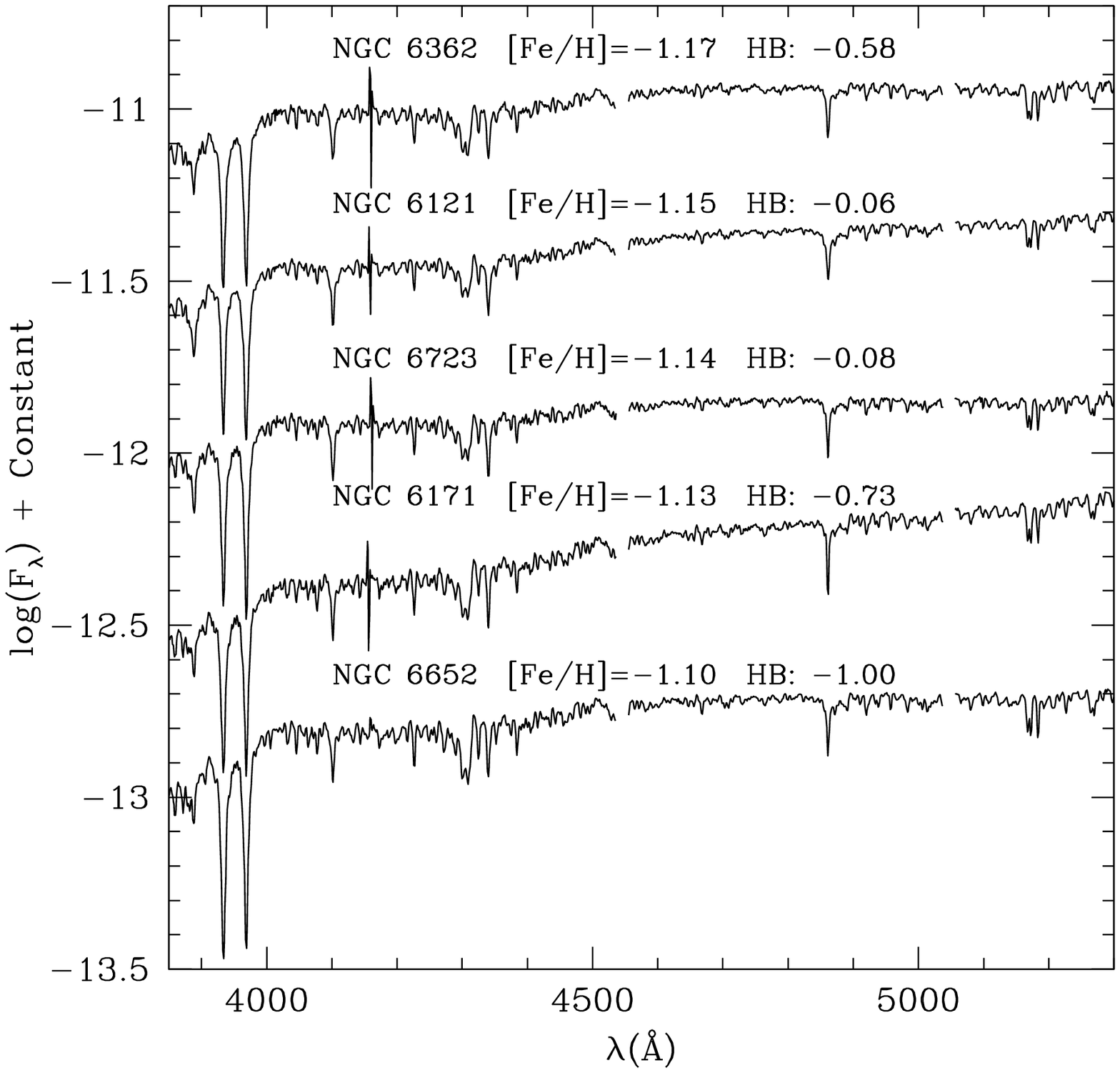}
\caption{e. Cont.}
\label{fig:spectra5}
\end{figure}
\setcounter{figure}{1}
\begin{figure}
\plotone{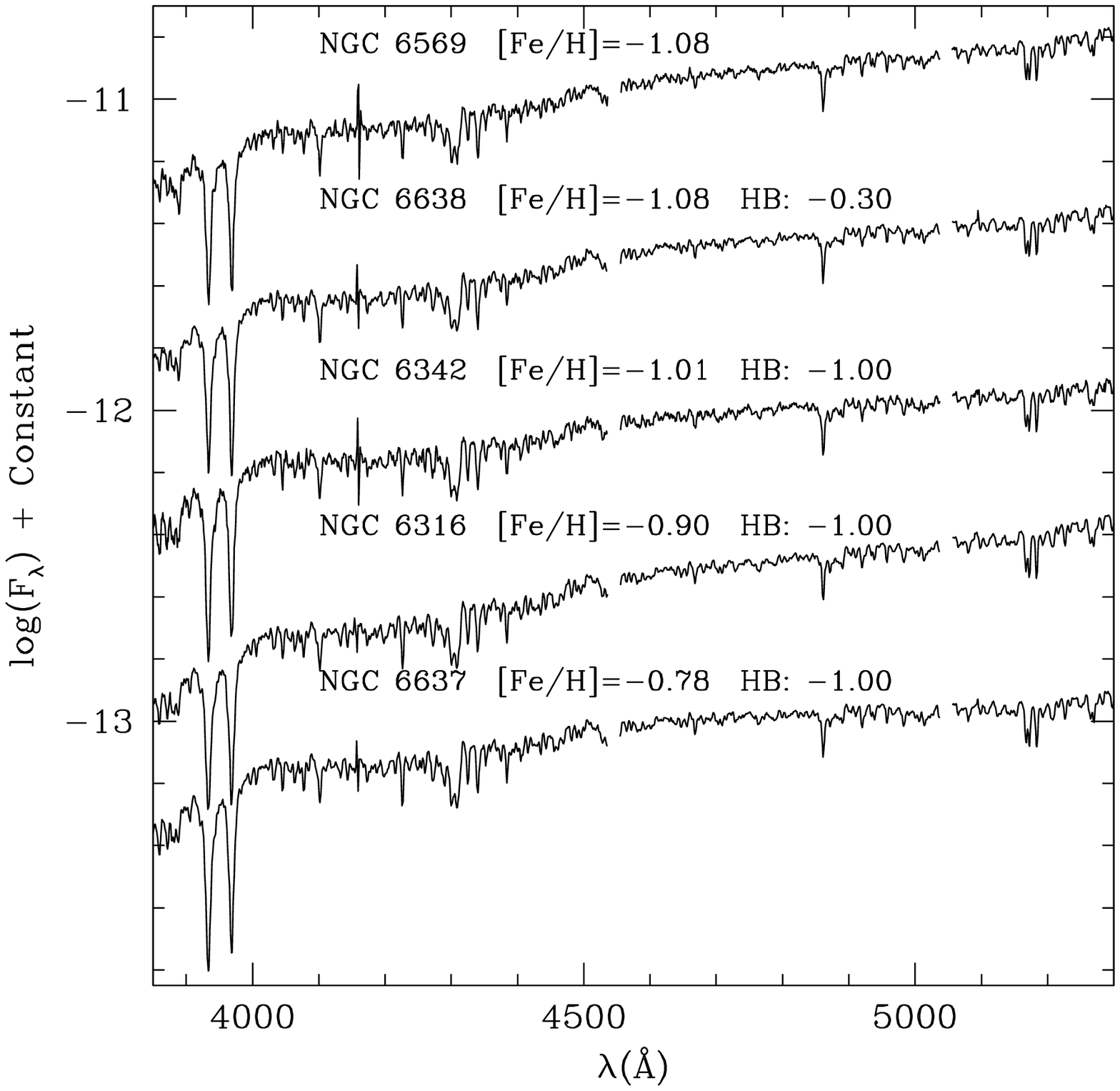}
\caption{f. Cont.}
\label{fig:spectra6}
\end{figure}
\setcounter{figure}{1}
\begin{figure}
\plotone{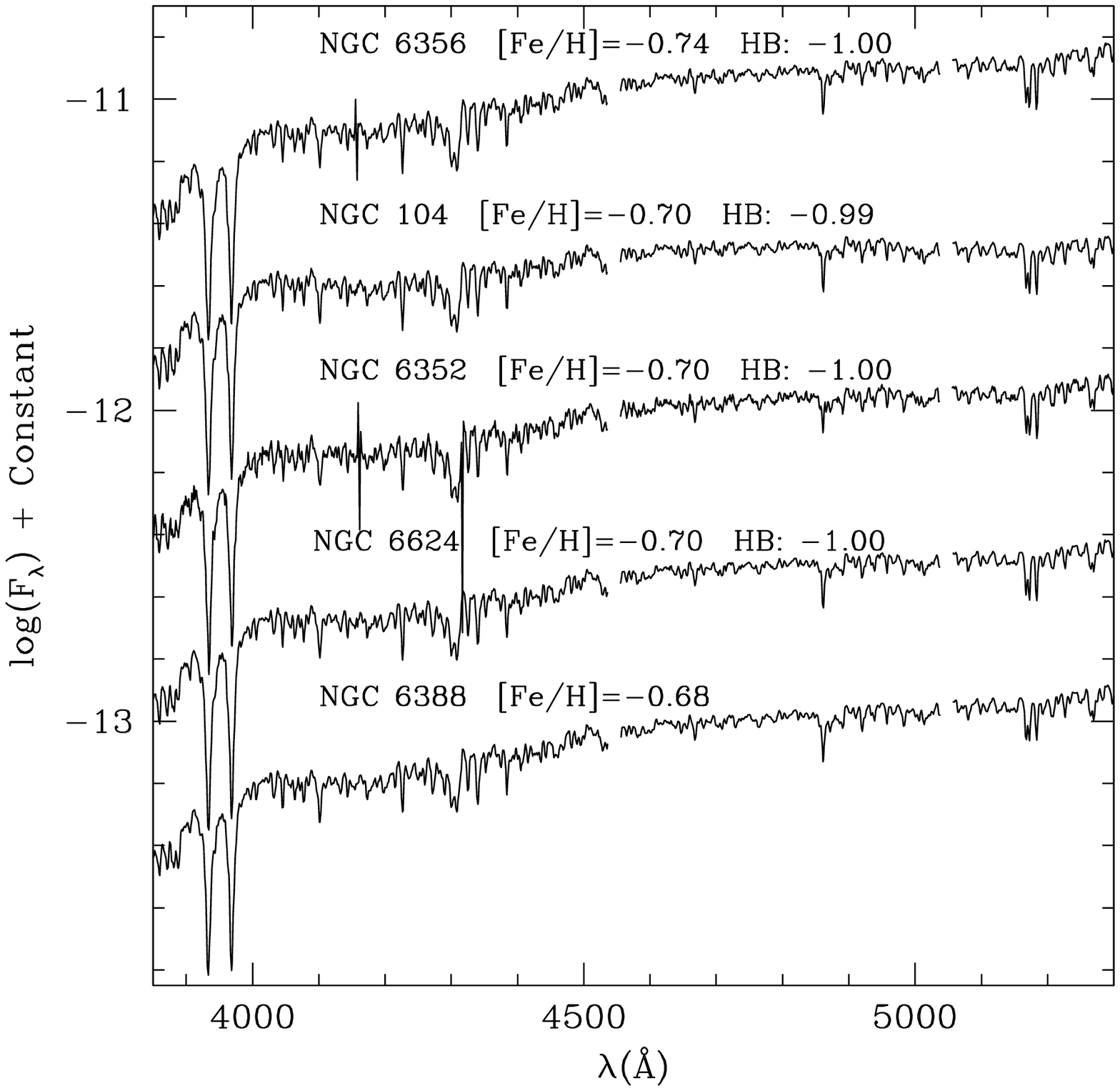}
\caption{g. Cont.}
\label{fig:spectra7}
\end{figure}
\setcounter{figure}{1}
\begin{figure}
\plotone{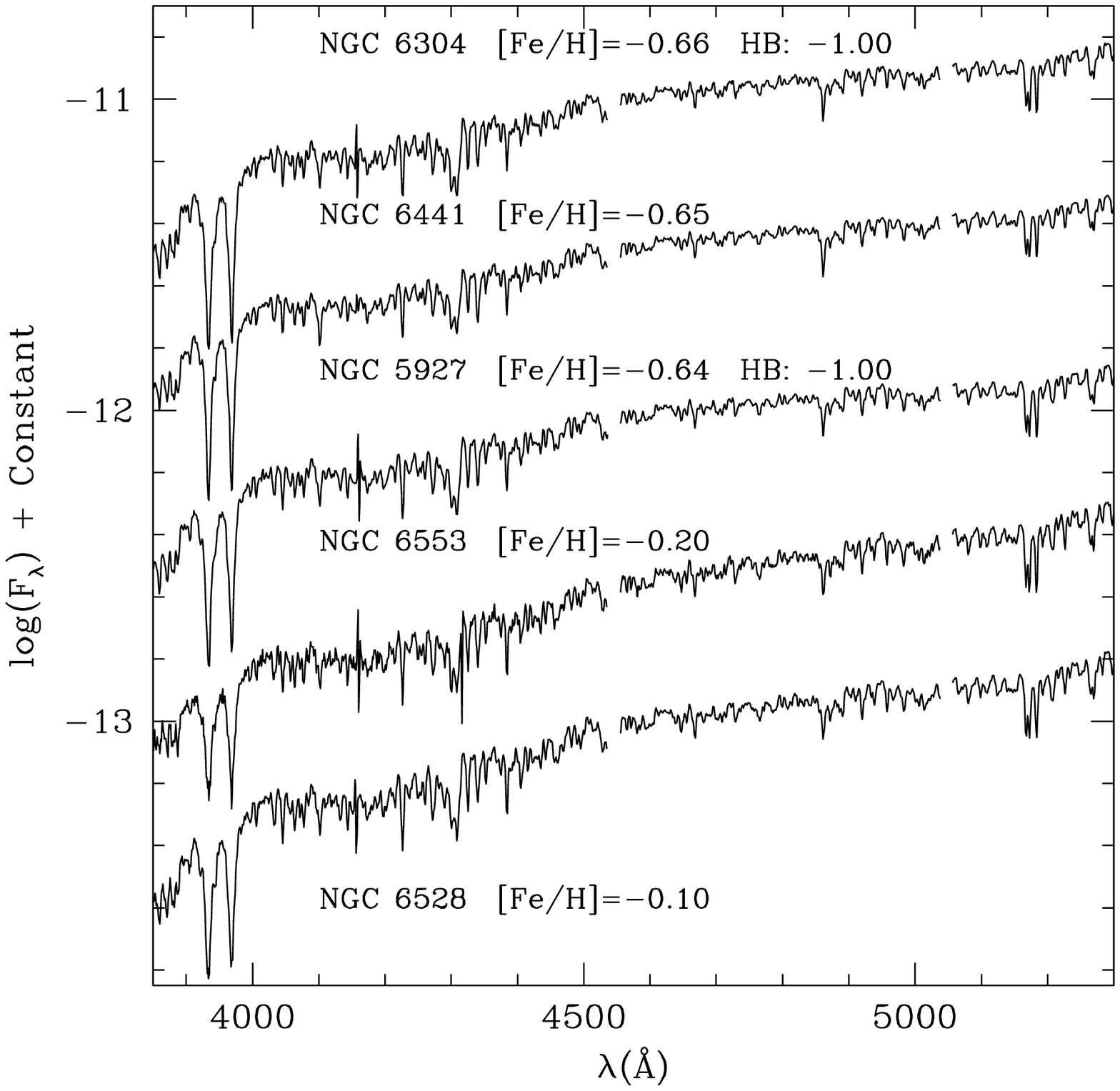}
\caption{h. Cont.}
\label{fig:spectra8}
\end{figure}

\clearpage

\begin{figure}
\plotone{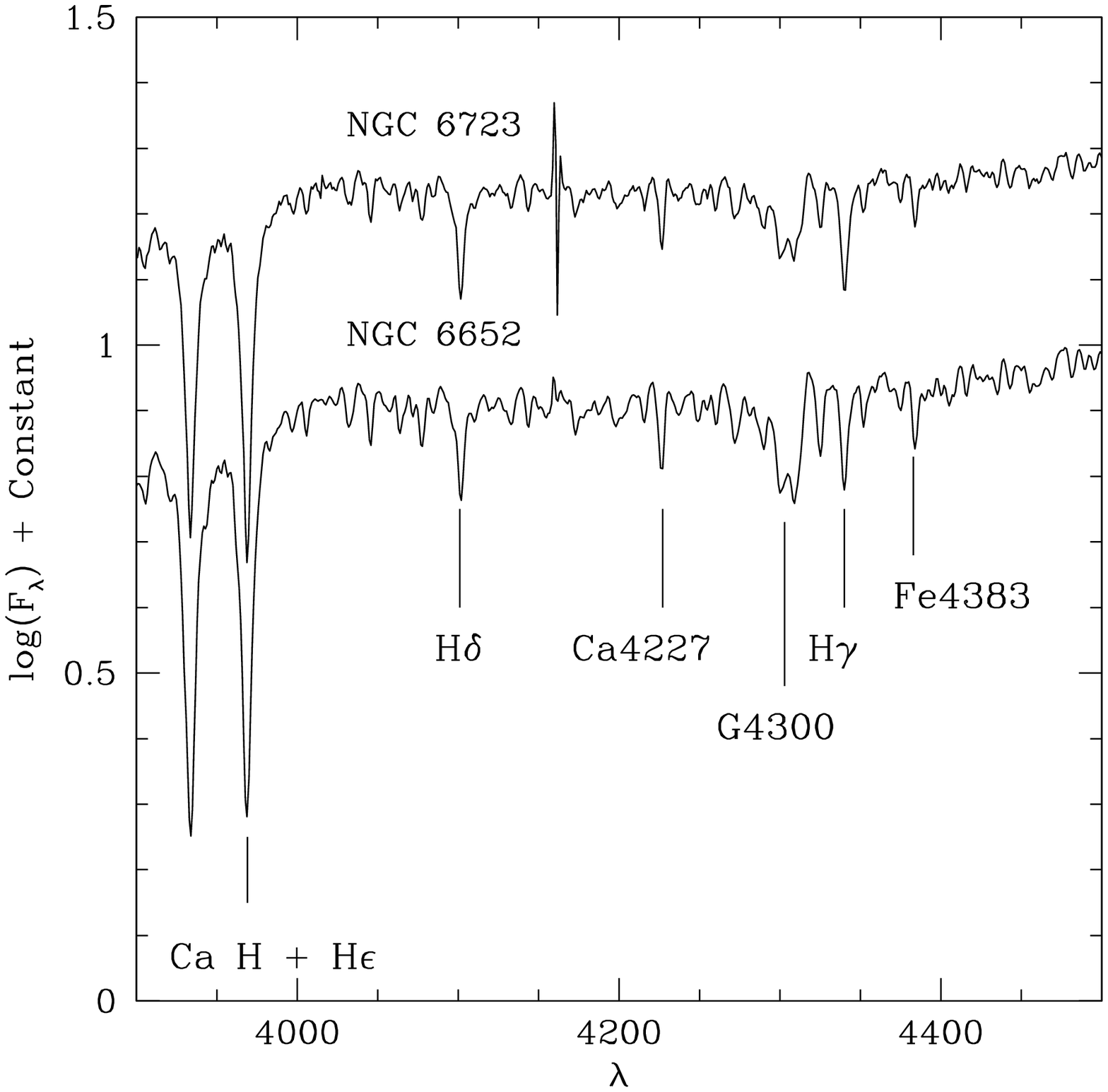}
\caption{Comparison of the integrated spectra of two moderately metal-poor
clusters ([Fe/H]$\sim$--1.1) and widely different horizontal-branch
morphologies. NGC 6652 has a strictly red horizontal branch, whereas NGC
6723 has a sizeable population of blue HB stars. As a result, Balmer lines
are stronger and metal lines are weaker in the spectrum of the latter.}
\label{fig:hbm}
\end{figure}

\begin{figure}
\plotone{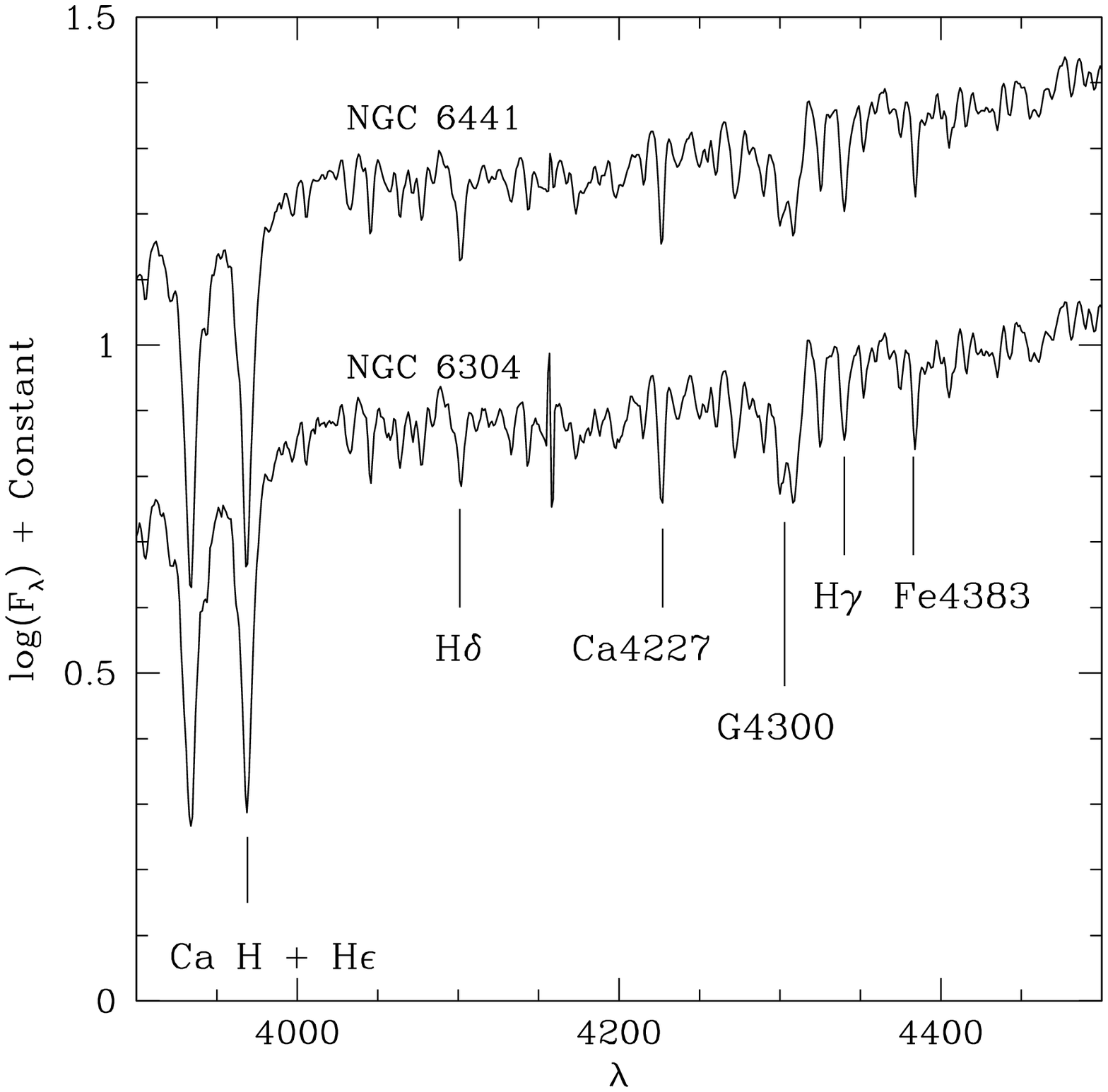}
\caption{Same as Figure~\ref{fig:hbm}, now for two metal-rich clusters. While
NGC 6304 has a stubby red horizontal branch, that of NGC 6441 is dominated
by red stars, but contains a sizeable blue extension. The effect here is
more subtle, in that only the Balmer lines are affected, being stronger
in the spectrum of NGC~6441. The metal lines are less affected, probably
because the horizontal-branch of NGC~6441 has a very strong red component.}
\label{fig:hbp}
\end{figure}

\begin{figure}
\plotone{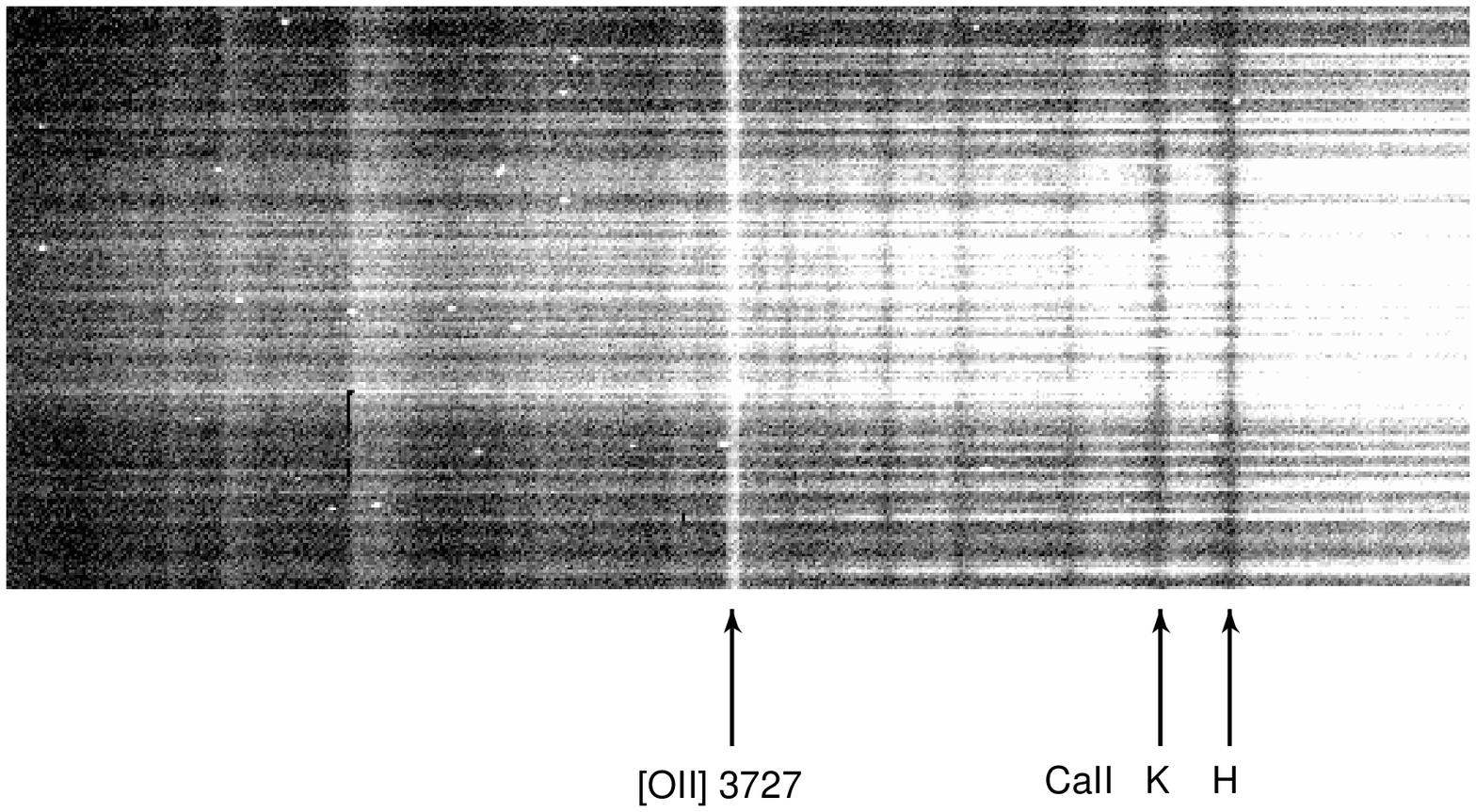}
\caption{Two-dimensional spectrum of NGC~3201. The presence of foreground
diffuse emission is indicated by the [\ion{O}{2}] 3727 line, suggesting
that Balmer lines in the cluster spectrum might be contaminated by
diffuse emission.  Careful background subtraction needs to be performed
in order to correctly remove this contamination.}
\label{fig:N3201}
\end{figure}

\clearpage

\begin{figure}
\plotone{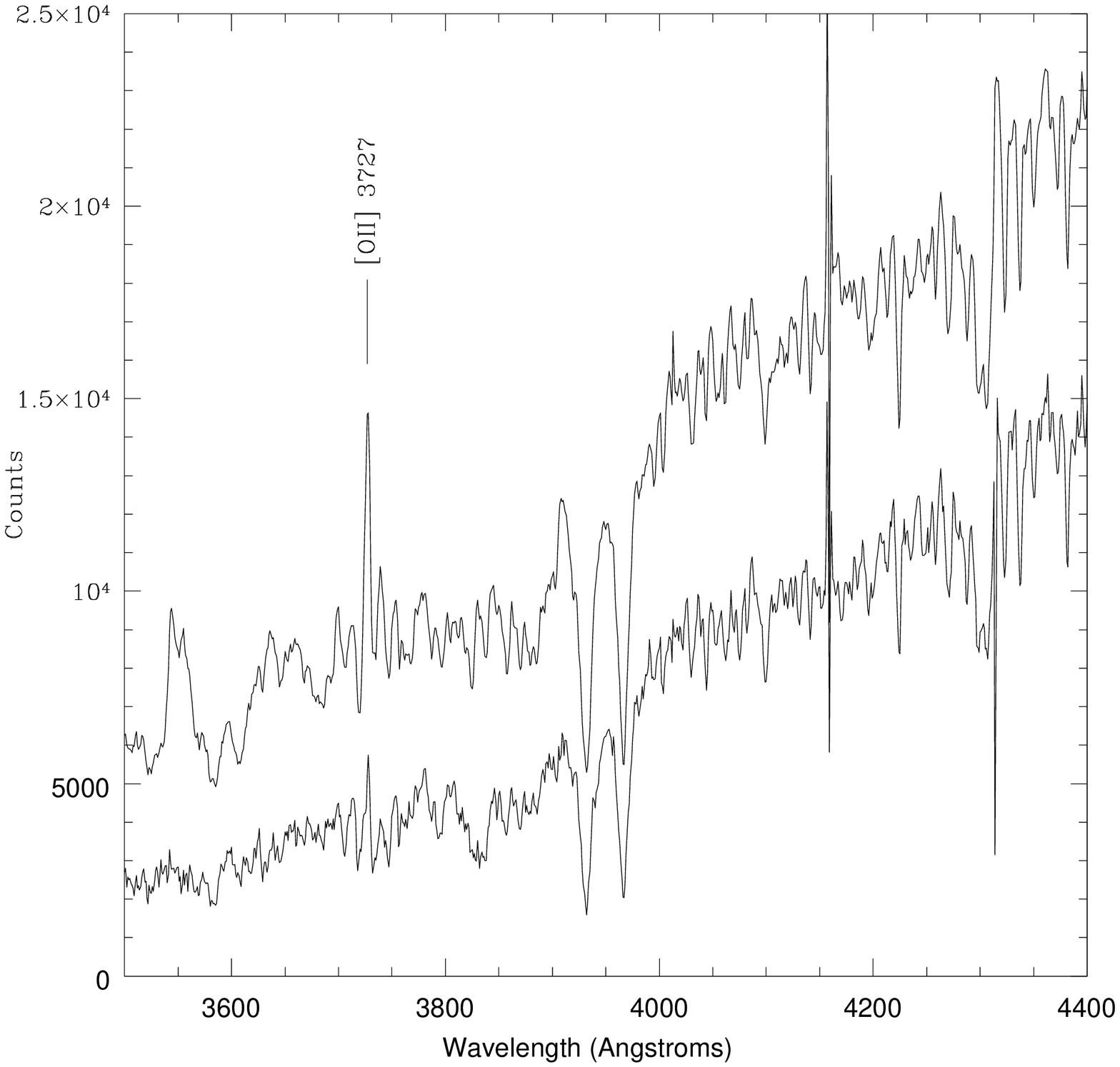}
\caption{Another example of contamination by diffuse foreground 
emission. Plotted are the spectrum of NGC 6352 before (top) and
after background subtraction. Note the very strong [\ion{O}{2}]
emission line in the top spectrum.}
\label{fig:n6352spec}
\end{figure}

\begin{figure}
\plotone{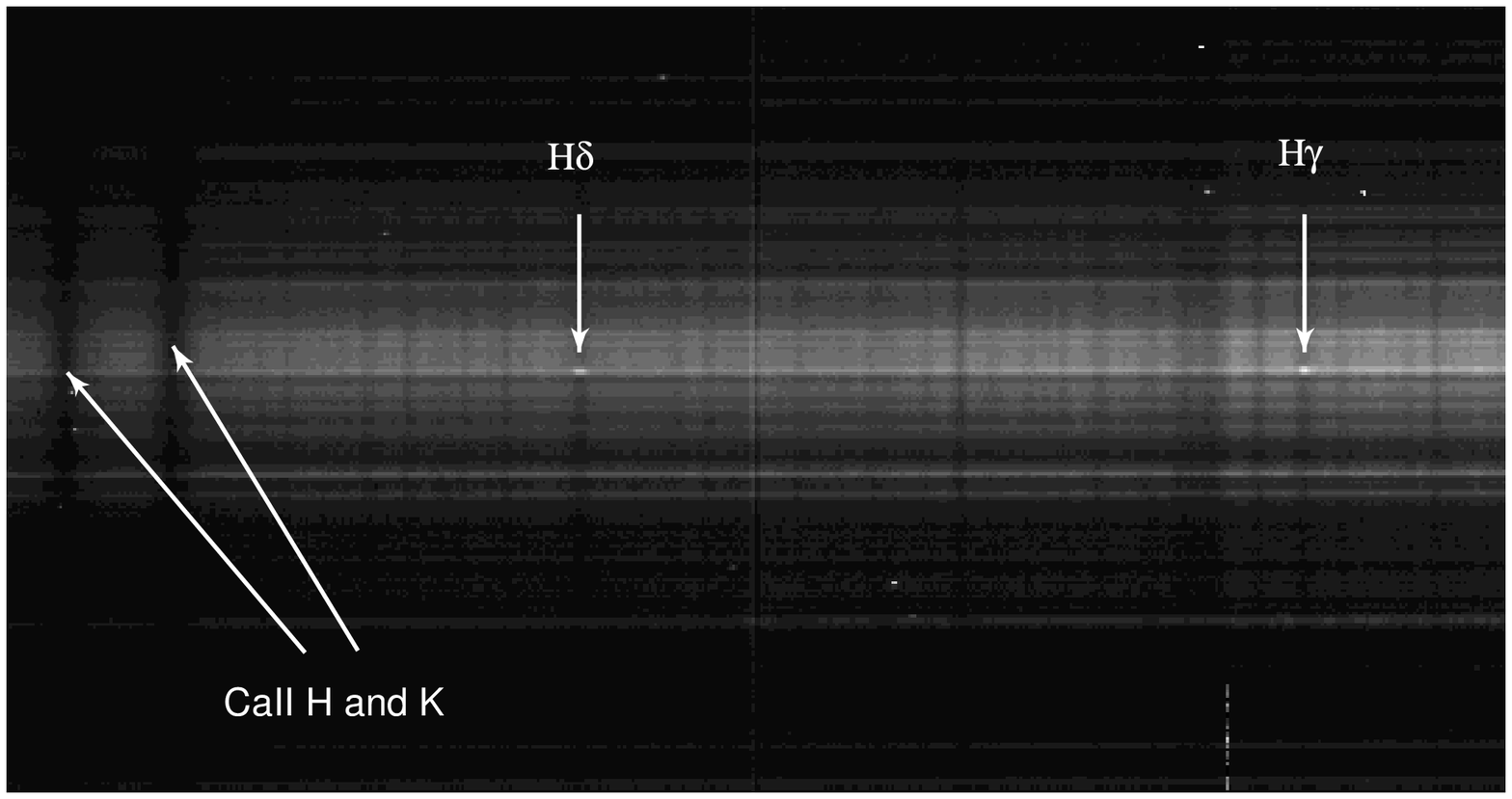}
\caption{Two-dimensional spectrum of NGC~6637. The arrows indicate
the presence of emission in both $H\delta$ and $H\gamma$ in a 
LPV star that was going through an emission-line phase during our
observing run.}
\label{fig:N6637im}
\end{figure}

\clearpage

\begin{figure}
\plotone{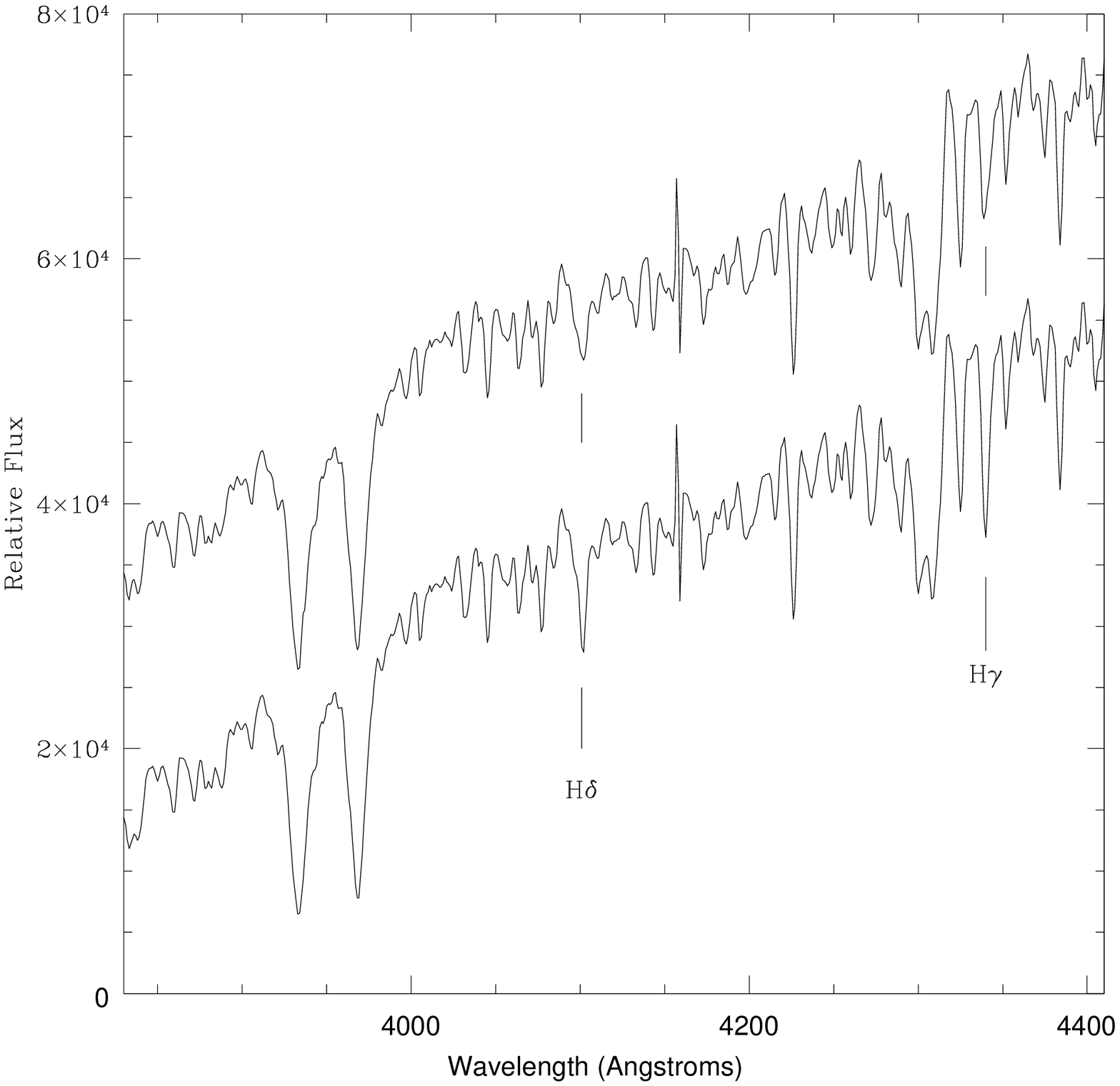}
\caption{One-dimensional spectra of NGC~6637 extracted from the 2-D
spectrum shown in Figure~\ref{fig:N6637im}. In the top spectrum, all
the light within the core radius was extracted, and in the bottom
spectrum, the LPV star indicated in Figure~\ref{fig:N6637im} was
excluded from the extraction window. Note the striking difference
in Balmer line strengths. Contamination by emission from the LPV
makes the Balmer lines much weaker in the top spectrum.}
\label{fig:N6637spec}
\end{figure}

\end{document}